\documentclass[lettersize,journal]{IEEEtran}

\usepackage{amsmath,amsfonts,amssymb}
\usepackage{array}
\usepackage[caption=false,font=normalsize,labelfont=sf,textfont=sf]{subfig}
\usepackage{textcomp}
\usepackage{stfloats}
\usepackage{url}
\usepackage{verbatim}
\usepackage{graphicx}
\usepackage{cite}

\usepackage{algorithm}
\usepackage{algpseudocode}

\usepackage{tabularray}
\UseTblrLibrary{booktabs}
\usepackage{bm}

\usepackage{hyperref}
\usepackage{orcidlink}  
\usepackage[capitalize,nameinlink]{cleveref}

\begin{document}

\title{Linking Dispersive-Medium Uncertainty to Clutter Analysis in Single-Snapshot FDA-MIMO-GPR}

\author{Yisu Yan$^{\orcidlink{0009-0008-4022-3619}}$,~\IEEEmembership{Graduate Student Member,~IEEE,} Jifeng Guo$^{\orcidlink{0000-0002-9710-0045}}$ 

	\thanks{Yisu Yan is with School of Astronautics, Harbin Institute of Technology, Heilongjiang, China
		(e-mail: \href{mailto:23B918085@stu.hit.edu.cn}{23B918085@stu.hit.edu.cn}).}

	\thanks{Jifeng Guo is with School of Astronautics, Harbin Institute of Technology, Heilongjiang, China (e-mail: \href{mailto:guojifeng@hit.edu.cn}{guojifeng@hit.edu.cn}).}

	\thanks{This work has been submitted to the IEEE for possible publication. Copyright may be transferred without notice, after which this version may no longer be accessible.}
}

\maketitle

\begin{abstract}
	Single-snapshot FDA-MIMO-GPR requires clutter models that account for dispersive-medium uncertainty, yet the statistical link between complex-medium characterization and clutter covariance analysis has remained unclear. This paper develops a propagation-side statistical framework that maps random perturbations of the relaxation spectrum to complex permittivity, complex wavenumber, steering-vector perturbation, medium-induced clutter covariance, and total clutter covariance. Within this framework, the effects of medium uncertainty on effective rank, effective clutter-subspace dimension, and target--clutter separability are characterized through a KL-based modal decomposition and a subspace-projection analysis. Numerical validation uses five literature-informed dielectric families to define physically traceable prior scenarios, a controlled random-field model to exercise the main propagation chain, and gprMax-based full-wave FDTD snapshots for an independent solver-level consistency check. Monte Carlo closure shows stage-wise numerical consistency, identifies steering linearization as the dominant approximation-sensitive step, and supports a weak perturbation regime with a bounded extension into a moderate regime. In a representative whitening-and-detection benchmark, the structured covariance model raises AUC from 0.593 for a diagonal baseline to 0.753, while prior-mismatch experiments indicate gradual rather than abrupt degradation. These results provide an explicit and interpretable interface for embedding complex-medium uncertainty into FDA-MIMO-GPR clutter analysis within a first-order, propagation-dominated setting.
\end{abstract}


\section{Introduction}\label{sec:intro}
\IEEEPARstart{T}{he} demand for subsurface sensing is shifting from static inspection to the real-time monitoring of dynamic processes, including tunnel-lining assessment \cite{wu2025GPRImageryBasedRealTime}, subsurface hydrological-fluid migration \cite{wen2024EnhancingImageAlignment}, and underground fingerprint matching for extraterrestrial exploration \cite{sheppard2025MarsLGPRMarsRover}. Ground-penetrating radar (GPR) is a promising platform for this shift because it is non-destructive and already supports dynamic sensing modes such as time-lapse GPR and Doppler GPR\cite{haruzi2020PotentialTimelapseGPR,johnson2007InsightsUseTimelapse,ruols20254DGPRImaging}. When coupled with signal-processing tools such as SVD and MUSIC\cite{yu2017DopplerRadarVital,yilmaz2025MusicGuidedSVD,liu2022VitalSignsFast,daluom2018TrackingMovingObject}, these systems can detect and track evolving subsurface targets.

For such scenarios, frequency diverse array (FDA) and multiple-input multiple-output (MIMO) architectures are appealing because they can acquire range--angle information synchronously under \emph{single-snapshot or few-snapshot} conditions, unlike the single-channel scanning mode common in conventional SFCW or FMCW systems. FDA introduces an additional range degree of freedom within one snapshot and can, in principle, focus energy at specific subsurface locations\cite{wang2023FrequencyDiverseArray}, while MIMO array designs such as Vivaldi-based configurations have also been explored for subsurface sensing in extraterrestrial settings\cite{liu2023DataPreProcessingSignal}. Yet FDA-MIMO-GPR faces a much harsher propagation environment than conventional airborne FDA-MIMO radar: near-field propagation, dispersion, attenuation, and uncertainty in medium parameters jointly reshape the space-frequency observation structure. As a result, clutter statistics, and their consequences for target--clutter separation, become central to single-snapshot FDA-MIMO-GPR analysis.

Existing FDA-MIMO clutter studies provide mature statistical tools but only a partial answer to this problem. Building on classical array-radar theory, prior work has developed covariance-, subspace-, and rank-based analyses\cite{goodman2007ClutterRankObserved,gini2002VectorSubspaceDetection} and then extended them to frequency-diverse settings, including clutter-rank characterization, range-ambiguous clutter analysis, joint space-time-range suppression, and parameter-design questions\cite{liu2016ClutterRanksFrequency,wang2022ClutterRankAnalysis,wen2019ClutterSuppressionAirborne,sun2024SpaceTimeRange,jia2026FDAMIMORadarParameter}. Related studies have further addressed clutter-covariance estimation, parameter estimation, and detection in complex clutter environments\cite{jain2023RadarClutterCovariance,rojhani2022CRLBsLocationVelocity,tian2022OutlierrobustTruncatedMaximum,xie2022RegularizedCovarianceEstimation,yang2022GroundClutterMitigation,zhang2021SealandClutterClassification}. However, this literature mainly attributes clutter structure to array configuration, platform motion, range ambiguity, and space-time coupling, while treating the medium itself in an oversimplified way\cite{liu2016ClutterRanksFrequency,wang2022ClutterRankAnalysis,wen2019ClutterSuppressionAirborne,sun2024SpaceTimeRange,jia2026FDAMIMORadarParameter}.

A complementary body of work in GPR and applied geophysics has long studied the missing physical ingredient: complex subsurface media. Complex permittivity has been used to describe attenuation and frequency-dependent dispersion in a unified way \cite{bano1996ConstantDielectricLosses}; Debye, Cole--Cole, and generalized relaxation models have been used to capture polarization mechanisms and time--frequency responses \cite{holm2020TimeDomainCharacterization}; and more recent studies have pursued relaxation-time-distribution estimation, statistical learning-based representations, arbitrary-permittivity electromagnetic modeling, and random-media simulation \cite{florsch2012DirectEstimationDistribution,liu2019GaussianProcessDistribution,majchrowska2021ModellingArbitraryComplex,jiang2013SimulationAnalysisGPR,mitLincolnLabLGPRTechNotes}. At the same time, the practical importance of clutter in GPR is clear from the large literature on clutter suppression, noise reduction, target detection, image enhancement, and data reconstruction \cite{bi2018ClutterEliminationRandomnoise,xue2019NoiseSuppressionGPR,kumlu2020GPRClutterReduction,oliveira2021GPRClutterReflection,su2022GPRImageClutter,liu2023GPRClutterRemoval,zhao2023ClutterRemovalMethod,dai2023GPRDataReconstruction,hoarau2017RobustAdaptiveDetection,worthmann2021ClutterDistributionsTomographic,wang2025AdvancedAdaptiveMedian}. What remains missing is an explicit statistical interface between these medium models and FDA-MIMO clutter analysis.

This paper develops that interface for single-snapshot FDA-MIMO-GPR. Starting from a random perturbation field on the logarithmic relaxation spectrum, medium uncertainty is propagated to the complex permittivity, complex wavenumber, steering-vector perturbation, medium-induced clutter covariance, and total clutter covariance. These covariance changes are then related to spectral broadening, effective rank, effective clutter-subspace dimension, and target--clutter separability. In this sense, the main contribution is an explicit and computable bridge between complex-medium uncertainty and FDA-MIMO clutter statistics. The analysis is primarily theoretical and numerical, spanning two tiers of validation: literature-informed dielectric families define physically traceable prior scenarios and a controlled Mat\'ern-based relaxation-spectrum perturbation model exercises the full propagation chain, while gprMax-based full-wave FDTD snapshots provide an independent solver-level consistency check. The regime in which the first-order characterization remains accurate is identified, and the resulting covariance is assessed in a controlled whitening-and-matched-filter detection benchmark. This benchmark is intended as a mechanism-level probe rather than a full end-to-end system evaluation.

Section \ref{sec:signal_model} presents the signal model and medium characterization. Section \ref{sec:theory} develops the propagation from medium randomness to steering perturbation and clutter covariance, and then analyzes the resulting spectral structure, effective rank, and separability. Section \ref{sec:experiments} validates the theory numerically and examines its effectiveness and applicability boundaries.

\section{FDA-MIMO GPR Signal Model}\label{sec:signal_model}

\subsection{General Model}\label{subsec:general_model}

A one-dimensional collocated FDA-MIMO GPR array with $M$ transmit--receive channels is considered, where the equivalent spatial position of the $m$th channel is denoted by $d_m$. The carrier frequency of the $m$th channel is denoted by $f_m$ and is written as
\begin{equation}\label{eq:f_m}
	f_m = f_0 + n_m \Delta f
\end{equation}
where $f_0$ is the reference center frequency, $\Delta f$ is the fundamental frequency increment, and $n_m$ is the frequency-coding coefficient.

Consider a single scattering element located at the parametric coordinate $(\theta,r)$, where $\theta$ denotes the azimuth angle and $r$ denotes the radial distance relative to the reference phase center. The local region containing this scatterer is assumed isotropic, with $\mu \approx \mu_0$, and no additional sources are assumed. The corresponding complex wavenumber therefore satisfies
\begin{equation}
	k_c^2(\omega;\bm{\mu})=\omega^2\mu_0\epsilon_0\epsilon_r^c(\omega;\bm{\mu})
\end{equation}
and hence
\begin{equation}\label{eq:k_c}
	k_c(\omega;\bm{\mu})=\omega\sqrt{\mu_0\epsilon_c(\omega)}
\end{equation}

For the $m$th channel, the baseband response of the scattering element is written as
\begin{equation}\label{eq:y_m}
	y_m(\theta,r)=\beta(\theta,r)\,G_m(\theta,r)\exp\!\big(-j k_c(\omega_m;\bm{\mu})L_m(\theta,r)\big)+w_m
\end{equation}
where $\beta(\theta,r)$ is the equivalent complex scattering coefficient, $\omega_m = 2\pi f_m$, $L_m(\theta,r)$ is the equivalent propagation path length for that channel, and $G_m(\theta,r)$ absorbs geometric spreading, antenna-pattern effects, and other slowly varying deterministic prefactors.

The analysis adopts a \emph{propagation-dominated conditional-scattering} model. Residual medium uncertainty is represented as a propagation-side perturbation entering the steering vector through the complex wavenumber $k_c(\omega;\bm{\mu})$, whereas the equivalent scattering coefficient $\beta(\theta,r)$ is treated as \emph{conditionally specified} after the reference background, local scatterer class, and local geometry have been fixed. This approximation isolates how random phase delay, attenuation, and space--frequency steering perturbations reshape the second-order clutter structure in a single snapshot. Its implications and limitations, including the additional covariance terms that would arise when $\beta(\theta,r)$ is medium-dependent, are discussed in Section~\ref{subsec:propagation_dominated_scope}.

Under this model, the space-frequency steering vector of the scattering element is given by
\begin{gather}\label{eq:steering}
	a_m(\theta,r;\bm\mu) = G_m(\theta,r)\exp\!\big(-j k_c(\omega_m;\bm\mu)L_m(\theta,r)\big) \\
	\bm a(\theta,r;\bm\mu) =
	\begin{bmatrix}
		a_1(\theta,r;\bm\mu) & a_2(\theta,r;\bm\mu) & \cdots & a_M(\theta,r;\bm\mu)
	\end{bmatrix}^{\top}
\end{gather}
Accordingly, the contribution of a single scattering element to one snapshot is
\begin{equation}\label{eq:snapshot_sg_patch}
	\bm y(\theta,r) = \beta(\theta,r)\bm a(\theta,r;\bm\mu) + \bm \omega
\end{equation}

When the scene contains a large number of distributed scattering elements, the received single-snapshot signal is represented as the superposition of local contributions over the scene. With residual terms such as system noise absorbed into $\bm n$, the continuous model becomes
\begin{equation}\label{eq:snapshot_all_patch}
	\bm x = \iint \beta(\theta,r)\bm a(\theta,r;\bm\mu)\,d\theta\,dr + \bm n
\end{equation}
Equation \eqref{eq:snapshot_all_patch} serves as the unified single-snapshot signal model for FDA-MIMO GPR. The subsequent clutter analysis is built on the scene-aggregated behavior of the steering-vector family $\{\bm a(\theta,r;\bm\mu)\}$.

\subsection{Characterization of Complex Media}\label{subsec:characterization_media}

As noted in Section~\ref{sec:intro}, GPR operates in subsurface media that exhibit conductive loss, dispersion, and spatial inhomogeneity. In the present single-snapshot space--frequency framework, these effects enter the observation model primarily through the complex permittivity and the associated complex wavenumber.

A continuous relaxation-spectrum representation is adopted as the primary medium model. Let $\tau$ denote the relaxation time constant, and define the logarithmic relaxation-time variable as
\begin{equation}\label{eq:log_tau}
	u
	=
	\log \tau
\end{equation}
The complex permittivity is then represented as
\begin{equation}\label{eq:epsilon_c}
	\epsilon_c(\omega)
	=
	\epsilon_{\infty}
	+
	\int_{-\infty}^{\infty}
	\frac{\tilde g(u)}{1+j\omega e^u}
	\,du
\end{equation}
where $\epsilon_{\infty}$ is the high-frequency limiting dielectric response and $\tilde g(u)$ is the relaxation-spectrum density on the logarithmic relaxation-time axis. Equation~\eqref{eq:epsilon_c} is taken as the starting point for the subsequent statistical analysis because it allows medium uncertainty to be represented directly as a perturbation of the relaxation spectrum.

The connection between \eqref{eq:epsilon_c} and classical relaxation models follows from the choice of $\tilde g(u)$. If the spectrum consists of a finite set of impulses,
\begin{equation}\label{eq:debye_discrete_spectrum}
	\tilde g(u)
	=
	\sum_{\ell=1}^{L}
	\Delta\epsilon_{\ell}
	\delta(u-u_{\ell})
\end{equation}
where $u_{\ell}=\log\tau_{\ell}$, substitution into \eqref{eq:epsilon_c} yields
\begin{equation}\label{eq:multipole_debye}
	\epsilon_c(\omega)
	=
	\epsilon_{\infty}
	+
	\sum_{\ell=1}^{L}
	\frac{\Delta\epsilon_{\ell}}{1+j\omega\tau_{\ell}}
\end{equation}
which is the multi-pole Debye model. A finite Debye expansion is therefore a discrete-spectrum special case of the continuous relaxation-spectrum model, and the single-pole Debye model is recovered by taking $L=1$.

For numerical implementation, the continuous spectrum in \eqref{eq:epsilon_c} can be discretized on a logarithmic relaxation-time grid $\{u_k\}_{k=1}^{K}$, giving
\begin{equation}\label{eq:continuous_to_discrete_debye}
	\epsilon_c(\omega)
	\approx
	\epsilon_{\infty}
	+
	\sum_{k=1}^{K}
	\frac{g_k\Delta u}{1+j\omega e^{u_k}}
\end{equation}
where $g_k\approx\tilde g(u_k)$. This discretized form is compatible with both finite Debye fitting and relaxation-time-distribution estimation.

The Cole--Cole model is interpreted within the same framework as a compact parameterization of broadened relaxation. In relative-permittivity notation, a conductive Cole--Cole response is commonly written as
\begin{equation}\label{eq:cole_cole_model}
	\epsilon_{r,c}^{\mathrm{CC}}(\omega)
	=
	\epsilon_{r,\infty}
	+
	\frac{\Delta\epsilon_r}
	{
		1+
		\left(
		j\omega\tau_0
		\right)^{1-\alpha}
	}
	+
	\frac{\sigma}{j\omega\epsilon_0}
\end{equation}
where $\tau_0$ is the characteristic relaxation time, $\alpha$ controls the relaxation broadening, and $\sigma$ represents conductive loss. The Debye response is recovered when $\alpha=0$, whereas $\alpha>0$ represents a broadened relaxation process that may be associated with a distribution of relaxation times. Cole--Cole parameters can therefore be used to generate or constrain a representative continuous relaxation spectrum over the operating band.

The continuous representation in \eqref{eq:epsilon_c} is also convenient for defining random perturbations and second-order statistics. Accordingly, the relaxation spectrum is decomposed as
\begin{equation}\label{eq:tilde_g_decomposition_intro}
	\tilde g(u)
	=
	\bar{\tilde g}(u)
	+
	\delta\tilde g(u)
\end{equation}
where $\bar{\tilde g}(u)$ denotes the nominal relaxation spectrum and $\delta\tilde g(u)$ denotes the random spectrum perturbation. This decomposition preserves compatibility with standard Debye and Cole--Cole descriptions while providing the statistical interface required for the covariance analysis developed below.

\section{Propagation of Medium Uncertainty to Clutter Statistics}\label{sec:theory}

This section develops the propagation of medium uncertainty from the relaxation-spectrum perturbation to the clutter covariance and then to the quantities that characterize spectral spreading, effective clutter-subspace dimension, and target--clutter separability:
\begin{gather*}
	\delta\tilde g(u)
	\rightarrow
	\delta\epsilon_c(\omega)
	\rightarrow
	\delta k_c(\omega)
	\rightarrow
	\delta\bm a(\theta,r) \\
	\rightarrow
	\bm R_a(\theta,r)
	\rightarrow
	\bm R_{\mathrm{med}}
	\rightarrow
	\bm R_c
	\rightarrow
	r_{\mathrm{eff}},\,p_\rho,\,\eta,\,\gamma
\end{gather*}

\subsection{Assumptions and Random Field Model}\label{subsec:assumptions_random_field}

The statistical assumptions required for the subsequent covariance propagation and spectral analysis are summarized first.

\newcounter{asmctr}
\renewcommand{\theasmctr}{A\arabic{asmctr}}

\begin{itemize}
	\refstepcounter{asmctr}
	\item[\textbf{\theasmctr}\label{asm:second_order}]
	      The logarithmic relaxation-spectrum perturbation $\delta\tilde g(u)$ is modeled as a zero-mean second-order random process:
	      \begin{equation}\label{eq:tilde_g}
		      \tilde g(u)=\bar{\tilde g}(u)+\delta \tilde g(u),\quad u=\log\tau
	      \end{equation}
	      with
	      \begin{equation}\label{eq:K}
		      \mathbb E[\delta\tilde g(u)]=0,\quad
		      \tilde K(u,u') = \mathbb E\left[\delta\tilde g(u)\delta\tilde g(u')^*\right]
	      \end{equation}
	      where $\tilde K(u,u')$ is a Hermitian positive semidefinite kernel.

	      \refstepcounter{asmctr}
	\item[\textbf{\theasmctr}\label{asm:technical_conditions}]
	      Standard regularity conditions are assumed for the integral operators used below, including integrability, square integrability, and interchangeability of expectation and integration. Thus, the Debye-type integral mappings, local covariance operators, and KL expansions are well defined.

	      \refstepcounter{asmctr}
	\item[\textbf{\theasmctr}\label{asm:propagation_dominated}\label{asm:independence_beta_da}]
	      A propagation-dominated conditional-scattering approximation is adopted. Residual medium uncertainty is represented by the steering perturbation $\delta\bm a(\theta,r)$, while $\beta(\theta,r)$ is treated as a conditionally specified scattering coefficient that is second-order independent of $\delta\bm a(\theta,r)$. Scattering-coefficient perturbations driven by the same medium randomness are excluded from the main covariance model; their effects are discussed in Section~\ref{subsec:propagation_dominated_scope}.

	      \refstepcounter{asmctr}
	\item[\textbf{\theasmctr}\label{asm:uncorrelated_scatterers}]
	      Distinct scattering elements are assumed to be second-order uncorrelated, with local power
	      \begin{equation}
		      \sigma_\beta^2(\theta,r)=\mathbb E[|\beta(\theta,r)|^2]
	      \end{equation}

	      \refstepcounter{asmctr}
	\item[\textbf{\theasmctr}\label{asm:first_order_phase}]
	      A weak propagation-phase perturbation condition is assumed for the first-order steering-vector expansion. The accumulated wavenumber perturbation along each propagation path is required to remain small in the mean-square sense, as quantified by
	      \begin{equation}\label{eq:rho_ph_def_assumption}
		      \rho_{\mathrm{ph}}
		      =
		      \max_{m,(\theta,r)\in\Omega}
		      \left\{
		      \mathbb E
		      \left[
			      \left|
			      L_m(\theta,r)\delta k_c(\omega_m)
			      \right|^2
			      \right]
		      \right\}^{1/2}
	      \end{equation}
	      where $\Omega$ denotes the clutter-support region used in the covariance construction. The first-order steering perturbation model is intended for the weak phase-perturbation regime $\rho_{\mathrm{ph}}\ll 1$. When $\rho_{\mathrm{ph}}$ approaches order unity, higher-order exponential phase terms may no longer be negligible.
\end{itemize}

As a baseline specification, $\delta\tilde g(u)$ is further modeled as a stationary process with a Matérn kernel:
\begin{equation}\label{eq:matern}
	\tilde K(u,u')=\sigma_g^2\kappa_{\nu,\ell}(u-u')
\end{equation}
where $\sigma_g^2$ determines the perturbation strength, while $\nu$ and $\ell$ control the smoothness and correlation scale, respectively.

\subsection{From Relaxation Spectrum to Steering Perturbation}\label{subsec:random_to_steering}

Under Assumption \ref{asm:second_order} and \ref{asm:technical_conditions}, substituting \eqref{eq:tilde_g} into \eqref{eq:epsilon_c} yields a decomposition of the complex permittivity
\begin{equation}\label{eq:epsilon_c_tilde_g}
	\epsilon_c(\omega) = \underbrace{\epsilon_\infty+\int_{-\infty}^{\infty}\frac{\bar{\tilde g}(u)}{1+j\omega e^u}\,du}_{\bar\epsilon_c(\omega)} + \underbrace{\int_{-\infty}^{\infty}\frac{\delta\tilde g(u)}{1+j\omega e^u}\,du}_{\delta\epsilon_c(\omega)}
\end{equation}
where $\bar{\epsilon}_c(\omega)$ is the nominal complex permittivity, and $\delta\epsilon_c(\omega)$ is a small perturbation satisfying
\begin{equation}\label{eq:delta_epsilon_c}
	\delta\epsilon_c(\omega) = \int_{-\infty}^{\infty}\frac{\delta\tilde g(u)}{1+j\omega e^u}\,du
	,\quad
	|\delta\epsilon_c(\omega)|\ll|\bar{\epsilon}_c(\omega)|
\end{equation}

The second-order statistics of $\epsilon_c(\omega)$ are therefore determined by the kernel $\tilde K$:
\begin{equation}\label{eq:cov_epsilon_c}
	\operatorname{Cov}\big(\epsilon_c(\omega),\epsilon_c(\omega')\big)
	=
	\int\int
	\frac{\tilde K(u,u')}{\big(1+j\omega e^u\big)\big(1-j\omega' e^{u'}\big)}
	\,du\,du'
\end{equation}

Taking the first-order expansion of \eqref{eq:k_c} about $\bar{\epsilon}_c(\omega)$ gives
\begin{equation}\label{eq:taylor_k_c}
	k_c(\omega;\bm{\mu})
	=
	k_c\big(\bar{\epsilon}_c(\omega)\big)
	+
	\left.\frac{\partial k_c(\omega;\bm{\mu})}{\partial \epsilon_c(\omega)}\right|_{\bar{\epsilon}_c(\omega)}
	\delta\epsilon_c(\omega)
	+
	o\big(\delta\epsilon_c(\omega)\big)
\end{equation}

Let $\bar{k}_c(\omega)=\omega\sqrt{\mu_0\bar{\epsilon}_c(\omega)}$. Then,
\begin{equation}\label{eq:delta_k_c}
	\begin{aligned}
		\delta k_c(\omega;\bm{\mu})
		 & = k_c(\omega;\bm{\mu})-\bar{k}_c(\omega)                                                  \\
		 & \approx \frac{\omega\mu_0}{2\sqrt{\mu_0\bar{\epsilon}_c(\omega)}}\delta\epsilon_c(\omega)
	\end{aligned}
\end{equation}
which can be further rewritten in terms of the relative permittivity. Let $\bar{\epsilon}_c(\omega)=\epsilon_0\bar{\epsilon}_r^c(\omega)$ and $\bar n_c(\omega)=\sqrt{\bar{\epsilon}_r^c(\omega)}$, then
\begin{equation}\label{eq:delta_k_c_n}
	\delta k_c(\omega;\bm{\mu}) \approx \frac{\omega}{2c_0\,\bar n_c(\omega)}\delta\epsilon_r^c(\omega)
\end{equation}

Combining this result with \eqref{eq:delta_epsilon_c} gives
\begin{equation}\label{eq:delta_k_c_tilde_g}
	\delta k_c(\omega;\bm{\mu})
	=
	\frac{\omega}{2c_0\bar n_c(\omega)}
	\int_{-\infty}^{\infty}\frac{\delta\tilde g(u)}{1+j\omega e^u}\,du
\end{equation}

For the $m$th channel, define the nominal steering component as
\begin{equation}\label{eq:a0m}
	a_{0,m}(\theta,r)
	=
	G_m(\theta,r)\exp\!\big(-j\bar{k}_c(\omega_m)L_m(\theta,r)\big)
\end{equation}

Then, according to \eqref{eq:steering}, one may write
\begin{equation}\label{eq:a_m_expand}
	a_m(\theta,r;\bm{\mu})
	=
	a_{0,m}(\theta,r)\exp\!\big(-j\,\delta k_c(\omega_m;\bm{\mu})L_m(\theta,r)\big)
\end{equation}

The accuracy of the following first-order expansion is governed by the accumulated propagation-phase perturbation, quantified by $\rho_{\mathrm{ph}}$ in \eqref{eq:rho_ph_def_assumption}. Its domain-averaged counterpart is
\begin{equation}\label{eq:rho_ph_bar_def}
	\bar{\rho}_{\mathrm{ph}}
	=
	\left\{
	\frac{1}{M|\Omega|}
	\sum_{m=1}^{M}
	\int_{\Omega}
	\mathbb E
	\left[
		\left|
		L_m(\theta,r)\delta k_c(\omega_m;\bm{\mu})
		\right|^2
		\right]
	d\theta dr
	\right\}^{1/2}
\end{equation}
where $M$ is the number of space--frequency channels. The index $\rho_{\mathrm{ph}}$ measures the largest mean-square accumulated phase perturbation over the channel--scene domain, whereas $\bar{\rho}_{\mathrm{ph}}$ provides the corresponding domain average. Both quantities are determined by the wavenumber-perturbation statistics, the operating frequencies, and the propagation geometry.

When $\rho_{\mathrm{ph}}\ll 1$, the exponential term in \eqref{eq:a_m_expand} admits a controlled first-order approximation. Therefore, under Assumption~\ref{asm:first_order_phase}, the steering perturbation is approximated as
\begin{equation}\label{eq:delta_a_m}
	\begin{aligned}
		\delta a_m(\theta,r)
		 & \triangleq
		a_m(\theta,r;\bm{\mu})-a_{0,m}(\theta,r) \\
		 & \approx
		-j\,a_{0,m}(\theta,r)\,L_m(\theta,r)\,\delta k_c(\omega_m;\bm{\mu})
	\end{aligned}
\end{equation}

This is the main linearization step in the propagation chain. The neglected terms scale with higher powers of $L_m(\theta,r)\delta k_c(\omega_m;\bm{\mu})$. The first-order steering approximation is controlled by both the spectrum perturbation and the accumulated phase perturbation. It can lose accuracy for moderate perturbations over long or strongly dispersive or attenuating paths.

Substituting \eqref{eq:delta_k_c_tilde_g} into \eqref{eq:delta_a_m} further gives
\begin{equation}\label{eq:delta_a_m_H}
	\delta a_m(\theta,r)
	\approx
	\int_{-\infty}^{\infty}
	H_m(\theta,r;u)\,\delta\tilde g(u)\,du
\end{equation}
where
\begin{equation}\label{eq:H_m}
	H_m(\theta,r;u)
	=
	-j\,L_m(\theta,r)\,a_{0,m}(\theta,r)\,
	\frac{\omega_m}{2c_0\bar n_c(\omega_m)}
	\frac{1}{1+j\omega_m e^u}
\end{equation}

Further define
\begin{equation}\label{eq:H_vec}
	\bm H(\theta,r;u)=\big[H_1(\theta,r;u),\dots,H_M(\theta,r;u)\big]^{\top}
\end{equation}

Then, the steering-vector perturbation can be expressed as
\begin{equation}\label{eq:delta_a_vec}
	\delta\bm a(\theta,r)
	=
	\int_{-\infty}^{\infty}
	\bm H(\theta,r;u)\,\delta\tilde g(u)\,du
\end{equation}

Medium randomness has been propagated from the relaxation-spectrum domain to the steering-vector domain, again, under first-order approximations valid under weak phase perturbations. Once the accumulated phase perturbation is no longer small, the nonlinear mapping in \eqref{eq:a_m_expand} can still be evaluated by Monte Carlo sampling, but the closed-form covariance propagation derived here is no longer quantitatively reliable.

\subsection{Local Steering Covariance and Scene-Level Covariance Decomposition}\label{subsec:covariance_decomposition}

Under Assumption \ref{asm:second_order} and \ref{asm:technical_conditions}, \eqref{eq:delta_a_vec} defines a well-posed second-order random vector and satisfies
\begin{equation}
	\mathbb E[\delta\bm a(\theta,r)] = \bm 0
\end{equation}

Define the local steering covariance as
\begin{equation}\label{eq:R_a_local}
	\bm R_a(\theta,r)
	\triangleq
	\mathbb E\!\left[\delta\bm a(\theta,r)\delta\bm a(\theta,r)^H\right]
\end{equation}

Then \eqref{eq:delta_a_vec}, \eqref{eq:H_vec}, Assumption \ref{asm:second_order}, and Assumption \ref{asm:technical_conditions} give
\begin{equation}\label{eq:R_a_local_integral}
	\bm R_a(\theta,r)
	=
	\int_{-\infty}^{\infty}
	\int_{-\infty}^{\infty}
	\bm H(\theta,r;u)\,
	\tilde K(u,u')\,
	\bm H(\theta,r;u')^H
	\,du\,du'
\end{equation}

Because $\tilde K$ is Hermitian positive semidefinite, $\bm R_a(\theta,r)$ is likewise Hermitian positive semidefinite.

Furthermore, the local steering vector can be decomposed as
\begin{equation}
	\bm a(\theta,r;\bm\mu)=\bm a_0(\theta,r)+\delta\bm a(\theta,r)
	\label{eq:decomposition_of_a}
\end{equation}
where
\begin{equation}
	\bm a_0(\theta,r)
	=
	\begin{bmatrix}
		a_{0,1}(\theta,r) & \cdots & a_{0,M}(\theta,r)
	\end{bmatrix}^{\top}
\end{equation}

Now consider the superposition of distributed scattering elements over the scene. From \eqref{eq:snapshot_all_patch}, the clutter snapshot can be written as
\begin{equation}
	\bm x_c
	=
	\iint
	\beta(\theta,r)\bm a(\theta,r;\bm\mu)\,d\theta\,dr
\end{equation}

Substituting \eqref{eq:decomposition_of_a} yields
\begin{equation}
	\bm x_c
	=
	\iint
	\beta(\theta,r)\bm a_0(\theta,r)\,d\theta\,dr
	+
	\iint
	\beta(\theta,r)\delta\bm a(\theta,r)\,d\theta\,dr
\end{equation}

Define the clutter covariance matrix as
\begin{equation}\label{eq:R_c_def}
	\bm R_c
	\triangleq
	\mathbb E[\bm x_c\bm x_c^H]
\end{equation}

Under Assumption~\ref{asm:propagation_dominated}, Assumption~\ref{asm:uncorrelated_scatterers}, and $\mathbb E[\delta\bm a(\theta,r)]=\bm 0$, the cross-covariance between the nominal term and the propagation-induced perturbation term vanishes in the main model, yielding
\begin{equation}\label{eq:R_c_decomp}
	\bm R_c=\bm R_0+\bm R_{\mathrm{prop}}
\end{equation}
where
\begin{equation}\label{eq:R_0}
	\bm R_0
	=
	\iint
	\sigma_\beta^2(\theta,r)\,
	\bm a_0(\theta,r)\bm a_0(\theta,r)^H
	\,d\theta\,dr
\end{equation}
denotes the clutter covariance under the nominal medium, while
\begin{equation}\label{eq:R_prop}
	\bm R_{\mathrm{prop}}
	=
	\iint
	\sigma_\beta^2(\theta,r)\,
	\bm R_a(\theta,r)
	\,d\theta\,dr
\end{equation}
represents the propagation-induced additional covariance term.

In the remainder of this paper, this term is denoted by
\begin{equation}\label{eq:R_med}
	\bm R_{\mathrm{med}}
	\equiv
	\bm R_{\mathrm{prop}}
\end{equation}
to emphasize that the medium-induced covariance considered here is the propagation-side component of the clutter covariance. Because $\bm R_a(\theta,r)$ is Hermitian positive semidefinite, both $\bm R_{\mathrm{med}}$ and $\bm R_c$ are also Hermitian positive semidefinite.

\subsection{Scope of the Propagation-Dominated Scattering Approximation}\label{subsec:propagation_dominated_scope}

The covariance construction above relies on a propagation-dominated approximation. In this approximation, the medium-dependent first-order randomness is assigned to the propagation operator: the random medium perturbs the complex wavenumber, path-dependent phase delay, attenuation, and FDA-MIMO space--frequency steering response. The equivalent scattering coefficient is treated as conditional on the reference background, scatterer class, local geometry, and scattering-strength statistics. This interpretation isolates the propagation-side component of medium-induced clutter covariance.

This distinction matters because practical GPR scattering coefficients may depend on the dielectric contrast between a scatterer and its surrounding background. If the same medium fluctuation changes both the propagation path and the local target--background contrast, the scattering coefficient should also be random. To make the consequence explicit, consider a discretized clutter cell indexed by $q$, whose local contribution is written as
\begin{equation}\label{eq:beta_random_local_response}
	\bm c_q
	=
	\beta_q\bm a_q
\end{equation}

Let
\begin{equation}\label{eq:beta_a_decomp}
	\begin{aligned}
		\beta_q
		 & =
		\bar\beta_q+\delta\beta_q \\
		\bm a_q
		 & =
		\bm a_{0,q}+\delta\bm a_q
	\end{aligned}
\end{equation}
where $\delta\beta_q$ denotes the scattering-coefficient perturbation and $\delta\bm a_q$ denotes the propagation-induced steering perturbation. Neglecting the second-order product $\delta\beta_q\delta\bm a_q$, one obtains
\begin{equation}\label{eq:local_response_first_order_beta}
	\bm c_q
	\approx
	\bar\beta_q\bm a_{0,q}
	+
	\bar\beta_q\delta\bm a_q
	+
	\delta\beta_q\bm a_{0,q}
\end{equation}

Accordingly, the clutter covariance generally contains four components,
\begin{equation}\label{eq:Rc_general_beta_random}
	\bm R_c
	\approx
	\bm R_0
	+
	\bm R_{\mathrm{prop}}
	+
	\bm R_{\beta}
	+
	\bm R_{\mathrm{cross}}
\end{equation}
where the propagation-induced term is
\begin{equation}\label{eq:Rprop_discrete}
	\bm R_{\mathrm{prop}}
	=
	\sum_q
	\mathbb E\!\left[
		|\bar\beta_q|^2
		\delta\bm a_q\delta\bm a_q^H
		\right]
\end{equation}
the scattering-coefficient perturbation term is
\begin{equation}\label{eq:Rbeta_discrete}
	\bm R_{\beta}
	=
	\sum_q
	\mathbb E\!\left[
		|\delta\beta_q|^2
		\bm a_{0,q}\bm a_{0,q}^H
		\right]
\end{equation}
and the cross term is
\begin{equation}\label{eq:Rcross_discrete}
	\begin{aligned}
		\bm R_{\mathrm{cross}}
		 & =
		\sum_q
		\mathbb E\!\left[
			\bar\beta_q\delta\bm a_q
			\left(\delta\beta_q\bm a_{0,q}\right)^H
			\right]
		\\ & +
		\sum_q
		\mathbb E\!\left[
			\delta\beta_q\bm a_{0,q}
			\left(\bar\beta_q\delta\bm a_q\right)^H
			\right]
	\end{aligned}
\end{equation}

The present analysis focuses on $\bm R_{\mathrm{prop}}$ in order to isolate the propagation-side statistical mechanism induced by dispersive background uncertainty. This restriction is most defensible in controlled or weakly coupled GPR settings, where medium variability mainly changes wave speed, attenuation, accumulated phase, and path-dependent channel response, while the equivalent scattering strength remains conditionally specified by scatterer type, geometry, or calibration statistics. Representative examples include dry or weakly moist granular soils, low-loss frozen ground, cold ice, dry snow, dry or frozen planetary-regolith analogs, and concrete or pavement inspection scenarios dominated by high-contrast objects, fixed layer interfaces, rocks, pipes, voids, rebars, or calibrated clutter cells.

The approximation can fail in strongly coupled environments. In wet clay, saline soils, water-rich or temperate ice, severely damaged concrete, or low-contrast dielectric-target scenarios, the same host-medium fluctuation may simultaneously change the propagation path and the local target--background contrast. In such cases, $\bm R_{\beta}$ and $\bm R_{\mathrm{cross}}$ should be modeled explicitly; their joint treatment is left for future work.

\subsection{Modal Interpretation of the Medium-Induced Covariance}\label{subsec:modal_interpretation}

Under Assumption \ref{asm:technical_conditions}, the kernel $\tilde K(u,u')$ admits the following Karhunen--Lo\`eve expansion:
\begin{equation}\label{eq:KL_kernel}
	\tilde K(u,u')
	=
	\sum_{q=1}^{\infty}
	\lambda_q
	\phi_q(u)\phi_q(u')^*
\end{equation}
where $\lambda_q\ge 0$ are the eigenvalues of the kernel, and $\{\phi_q\}$ are the corresponding orthonormal eigenfunctions. Substituting \eqref{eq:KL_kernel} into \eqref{eq:R_a_local_integral} gives
\begin{equation}\label{eq:R_a_KL}
	\bm R_a(\theta,r)
	=
	\sum_{q=1}^{\infty}
	\lambda_q\,
	\bm h_q(\theta,r)\bm h_q(\theta,r)^H
\end{equation}
where
\begin{equation}\label{eq:h_q}
	\bm h_q(\theta,r)
	\triangleq
	\int_{-\infty}^{\infty}
	\bm H(\theta,r;u)\phi_q(u)\,du
\end{equation}
denotes the image of the $q$th random relaxation-spectrum mode in the steering-vector domain.

Substituting this result further into \eqref{eq:R_med} yields
\begin{equation}\label{eq:R_med_KL}
	\bm R_{\mathrm{med}}
	=
	\sum_{q=1}^{\infty}
	\lambda_q\bm S_q
\end{equation}
where
\begin{equation}\label{eq:S_q}
	\bm S_q
	\triangleq
	\iint
	\sigma_\beta^2(\theta,r)\,
	\bm h_q(\theta,r)\bm h_q(\theta,r)^H
	\,d\theta\,dr
\end{equation}

Thus, $\bm R_{\mathrm{med}}$ is a weighted superposition of nonnegative modal components $\{\bm S_q\}$. The coefficients $\lambda_q$ characterize the energy distribution of medium randomness in the relaxation-spectrum domain, whereas $\bm S_q$ describes how the corresponding mode appears in the observation domain after propagation and scene-level integration.

\subsection{Spectral Broadening and Effective Rank}\label{subsec:spectral_broadening_rank}

The degree of covariance spectral broadening is characterized by the effective rank,
\begin{equation}\label{eq:r_eff_def}
	r_{\mathrm{eff}}(\bm R)
	\triangleq
	\frac{\big(\operatorname{tr}\bm R\big)^2}{\|\bm R\|_F^2}
\end{equation}
whose range satisfies $1\le r_{\mathrm{eff}}(\bm R)\le \operatorname{rank}(\bm R)$.

For the medium-induced covariance $\bm R_{\mathrm{med}}$, \eqref{eq:R_med_KL} gives
\begin{equation}\label{eq:r_eff_med_trace}
	\operatorname{tr}\bm R_{\mathrm{med}}
	=
	\sum_{q=1}^{\infty}
	\lambda_q\,\operatorname{tr}\bm S_q
\end{equation}
and
\begin{equation}\label{eq:r_eff_med_fro}
	\|\bm R_{\mathrm{med}}\|_F^2
	=
	\sum_{q=1}^{\infty}\sum_{p=1}^{\infty}
	\lambda_q\lambda_p
	\left\langle
	\bm S_q,\bm S_p
	\right\rangle_F
\end{equation}
where $\langle \bm A,\bm B\rangle_F=\operatorname{tr}(\bm A^H\bm B)$. Hence,
\begin{equation}\label{eq:r_eff_med}
	r_{\mathrm{eff}}(\bm R_{\mathrm{med}})
	=
	\frac{\left(\sum_q\lambda_q\operatorname{tr}\bm S_q\right)^2}
	{\sum_q\sum_p\lambda_q\lambda_p\langle\bm S_q,\bm S_p\rangle_F}
\end{equation}

This expression shows that the effective rank of $\bm R_{\mathrm{med}}$ depends not only on the distribution of the kernel spectrum $\{\lambda_q\}$, but also on the degree of geometric overlap among the modal components $\{\bm S_q\}$ in the channel domain.

For the total clutter covariance $\bm R_c=\bm R_0+\bm R_{\mathrm{med}}$, \eqref{eq:r_eff_def} yields
\begin{equation}\label{eq:r_eff_total}
	r_{\mathrm{eff}}(\bm R_c)
	=
	\frac{\big(\operatorname{tr}\bm R_0+\operatorname{tr}\bm R_{\mathrm{med}}\big)^2}
	{\|\bm R_0\|_F^2+\|\bm R_{\mathrm{med}}\|_F^2+2\langle \bm R_0,\bm R_{\mathrm{med}}\rangle_F}
\end{equation}

Further define
\begin{equation}\label{eq:mu_def}
	\mu
	\triangleq
	\frac{\langle \bm R_0,\bm R_{\mathrm{med}}\rangle_F}
	{\|\bm R_0\|_F\,\|\bm R_{\mathrm{med}}\|_F}
\end{equation}

Since both $\bm R_0$ and $\bm R_{\mathrm{med}}$ are Hermitian positive semidefinite matrices,
\begin{equation}
	\langle \bm R_0,\bm R_{\mathrm{med}}\rangle_F
	=
	\operatorname{tr}(\bm R_0\bm R_{\mathrm{med}})\ge 0
\end{equation}
it follows that $\mu\in[0,1]$. Accordingly, \eqref{eq:r_eff_total} can be rewritten as
\begin{equation}\label{eq:r_eff_total_mu}
	r_{\mathrm{eff}}(\bm R_c)
	=
	\frac{\big(\operatorname{tr}\bm R_0+\operatorname{tr}\bm R_{\mathrm{med}}\big)^2}
	{\|\bm R_0\|_F^2+\|\bm R_{\mathrm{med}}\|_F^2+2\mu\|\bm R_0\|_F\|\bm R_{\mathrm{med}}\|_F}
\end{equation}

The parameter $\mu$ quantifies the alignment between the nominal clutter covariance and the medium-induced covariance under the Frobenius inner product. A smaller $\mu$ implies a larger deviation of the additional covariance component from the original spectral structure and therefore a higher likelihood of increasing the total effective rank for the same trace increment.

Equations \eqref{eq:r_eff_med} and \eqref{eq:r_eff_total_mu} suggest that spectral broadening induced by medium randomness typically requires two conditions. First, the kernel spectrum $\{\lambda_q\}$ should not be overly concentrated on only a few modes. Second, the mapped modal components $\{\bm S_q\}$ should exhibit sufficient diversity in the channel domain. The former determines the effective degrees of freedom introduced by medium randomness, whereas the latter determines whether those degrees of freedom produce resolvable spectral expansion in the observation domain.

\subsection{Effective Clutter Subspace Dimension and Separability}\label{subsec:subspace_separability}

Consider the eigendecomposition of the total clutter covariance $\bm R_c$:
\begin{equation}
	\bm R_c
	=
	\sum_{m=1}^{M}
	\lambda_m^{(c)}
	\bm u_m\bm u_m^H,
	\quad
	\lambda_1^{(c)}\ge \lambda_2^{(c)}\ge \cdots \ge \lambda_M^{(c)}\ge 0
\end{equation}

For a prescribed energy threshold $\rho\in(0,1)$, define the effective clutter-subspace dimension as
\begin{equation}\label{eq:p_rho}
	p_\rho
	\triangleq
	\min\left\{
	p:\;
	\frac{\sum_{m=1}^{p}\lambda_m^{(c)}}{\sum_{m=1}^{M}\lambda_m^{(c)}}
	\ge \rho
	\right\}
\end{equation}

Accordingly, the subspace spanned by the first $p_\rho$ eigenvectors,
\begin{equation}
	\mathcal U_c^{(\rho)}
	=
	\operatorname{span}\{\bm u_1,\dots,\bm u_{p_\rho}\}
	\label{eq:U_c}
\end{equation}
can be regarded as the effective clutter subspace in the energy sense.

Let
\begin{equation}
	p_m=\frac{\lambda_m^{(c)}}{\sum_{i=1}^{M}\lambda_i^{(c)}},
	\quad
	\sum_{m=1}^{M}p_m=1
\end{equation}

Then,
\begin{equation}
	r_{\mathrm{eff}}(\bm R_c)=\frac{1}{\sum_{m=1}^{M}p_m^2}
\end{equation}

Moreover, from \eqref{eq:p_rho}, it follows that $\sum_{m=1}^{p_\rho}p_m\ge \rho$. By the Cauchy--Schwarz inequality,
\begin{equation}
	\sum_{m=1}^{p_\rho}p_m^2
	\ge
	\frac{\left(\sum_{m=1}^{p_\rho}p_m\right)^2}{p_\rho}
	\ge
	\frac{\rho^2}{p_\rho}
\end{equation}
which yields the lower bound
\begin{equation}\label{eq:prho_lower_bound}
	p_\rho \ge \rho^2\, r_{\mathrm{eff}}(\bm R_c)
\end{equation}

This result indicates that, as the effective rank increases, the effective clutter-subspace dimension associated with a prescribed energy threshold cannot remain arbitrarily small. Empirically, when the eigenspectrum is relatively smooth and no excessively dominant mode is present, $p_\rho$ often follows a monotonic trend consistent with $r_{\mathrm{eff}}(\bm R_c)$, although the two quantities are not identical in general.

Let $\bm a_t$ denote the target steering vector, and let the subspace spanned by the first $p$ dominant clutter eigenvectors, \eqref{eq:U_c} be taken as the principal clutter subspace. Its orthogonal projection matrix is
\begin{equation}
	\bm P_c^{(p)}
	=
	\sum_{m=1}^{p}\bm u_m\bm u_m^H
\end{equation}

Then, define the overlap between the target and the principal clutter subspace, and the corresponding separability measure, respectively, as
\begin{equation}\label{eq:gamma_p}
	\gamma(p)
	\triangleq
	\frac{\|\bm P_c^{(p)}\bm a_t\|_2^2}{\|\bm a_t\|_2^2}
\end{equation}
and
\begin{equation}\label{eq:eta_p}
	\eta(p)
	\triangleq
	\frac{\|(\bm I-\bm P_c^{(p)})\bm a_t\|_2^2}{\|\bm a_t\|_2^2}
	=
	1-\gamma(p)
\end{equation}

A larger $\gamma(p)$ indicates that a greater portion of the target energy falls within the principal clutter subspace, whereas a larger $\eta(p)$ implies improved separability between the target and the clutter subspace.

Accordingly, when medium randomness increases and causes both the effective rank of $\bm R_c$ and $p_\rho$ to grow, the principal clutter subspace generally occupies a broader portion of the observation space. For a fixed target steering vector $\bm a_t$, this typically increases $\gamma(p_\rho)$ and decreases $\eta(p_\rho)$, implying degraded target--clutter separability. In this sense, the effect of uncertainty in dispersive media on detection performance can be traced to its expansion of the clutter-covariance spectral structure and of the effective clutter subspace dimension.

\section{Numerical Validation}\label{sec:experiments}

\subsection{Numerical Setup}\label{sec:numerical_setup}

\subsubsection{Analytical Statistical Simulation Setup}
\label{sec:analytical_simulation_setup}

Five representative media (denoted S1--S5) anchor the experiments. They cover a low-loss dry-regolith baseline (S1), a strongly dispersive basalt or moist-medium anchor (S2), a frozen-ground or ice-like anchor (S3), and two engineering or fine-grained lossy media (S4--S5). The numerical configuration and medium parameters are reported in \cref{tbl:numerical_config,tbl:medium_parameters}. The baseline uses $M=16$ space--frequency channels, a logarithmic $200$--$269.3$ MHz FDA grid, $8 \times 8$ scene patches, $\sigma_g=0.03$, and $N_{\mathrm{MC}}=2000$.

Following the Cole--Cole parameter-family route, literature-informed dielectric parameters are sampled and propagated to complex permittivity and wavenumber, retaining the discrete relaxation-spectrum interpretation as the theoretical interface. This avoids an inverse problem in each Monte Carlo loop while preserving consistency with the continuous-spectrum formulation. The resulting families generate nonzero uncertainty in both $\epsilon_r$ and $k_c$, making the numerical prior a controlled dielectric family rather than an arbitrary perturbation source.

\begin{table}[!t]
	\centering
	\caption{Baseline numerical configuration used throughout the standalone validation package.}
	\label{tbl:numerical_config}
	\begin{tblr}{
		width=\columnwidth,
		colspec={X[1.5,l]X[1.0,c]X[1.8,l]},
		row{1}={font=\bfseries},
		cells={font=\footnotesize},
		}
		\toprule
		Parameter                                          & Value              & Role                                     \\
		\midrule
		Number of space--frequency channels $M$            & 16                 & FDA-MIMO observation dimension           \\
		Frequency grid                                     & $200$--$269.3$ MHz & Log-FDA operating band                   \\
		Scene patch grid                                   & $8 \times 8$       & Discrete $(\theta,r)$ aggregation grid   \\
		Relaxation-time grid size $K_u$                    & 256                & Discretized logarithmic spectrum support \\
		Baseline perturbation strength $\sigma_g$          & 0.03               & Reference random-field amplitude         \\
		Baseline Monte Carlo sample size $N_{\mathrm{MC}}$ & 2000               & Reference closure budget                 \\
		Main spectral threshold $\rho$                     & 0.9                & Effective subspace metric setting        \\
		Representative seed                                & 20260315           & Reproducible baseline realization        \\
		\bottomrule
	\end{tblr}
\end{table}

\begin{table*}[!t]
	\centering
	\caption{Literature-informed Cole--Cole or Debye-like medium parameters used to define the five representative scenarios.}
	\label{tbl:medium_parameters}
	\begin{tblr}{
		width=\textwidth,
		colspec={X[0.6,c]X[2.2,l]X[0.9,c]X[0.8,c]X[0.8,c]X[1.0,c]X[0.7,c]X[0.9,c]},
		row{1}={font=\bfseries},
		cells={font=\footnotesize},
		}
		\toprule
		Scene & Medium type                                                                                       & Model                   & $\epsilon_s$ & $\epsilon_\infty$ & $\tau$ (s)           & $\alpha$ & $\sigma$ (S/m)      \\
		\midrule
		S1    & Dry lunar regolith / dry low-loss soil analog \cite{strangway1974electricalLunarSoil}             & Cole--Cole              & 3.05         & 3.00              & $1.0\times10^{-6}$   & 0.30     & $1.0\times10^{-14}$ \\
		S2    & Basalt / moist-dispersive analog \cite{olhoeft1973electricalPropertiesLunarBasalt}                & Cole--Cole              & 1000         & 8.0               & $1.0\times10^{-6}$   & 0.30     & $1.0\times10^{-8}$  \\
		S3    & Water ice / frozen-ground anchor \cite{auty1952dielectricPropertiesIce}                           & Cole--Cole / Debye-like & 91.0         & 3.15              & $2.5\times10^{-5}$   & 0.00     & $1.0\times10^{-8}$  \\
		S4    & Water-bearing kaolinite / concrete-like anchor \cite{mansour2020dielectricKaoliniteWaterSediment} & Cole--Cole              & 35.6         & 2.0               & $5.0\times10^{-12}$  & 0.20     & 0.08                \\
		S5    & Fine-grained clay / pavement-soil anchor \cite{schwing2013dielectricClaySoil}                     & Cole--Cole              & 30.26        & 10.7              & $9.55\times10^{-12}$ & 0.062    & 0.0                 \\
		\bottomrule
	\end{tblr}
\end{table*}

\cref{tbl:medium_summary} summarizes the medium-family statistics extracted from the standalone data package. All five media produce nonzero uncertainty in both complex permittivity and complex wavenumber, and the averaged phase-perturbation index $\bar{\rho}_{\mathrm{ph}}$ varies substantially across scenarios, from $0.357$ for the low-loss baseline to $4.260$ for the most dispersive anchor.

\begin{table*}[!t]
	\centering
	\caption{Literature-informed medium-family summary.}
	\label{tbl:medium_summary}
	\begin{tblr}{
		width=\textwidth,
		colspec={X[0.8,c]X[2.4,l]X[1.2,c]X[1.2,c]X[1.2,c]},
		row{1}={font=\bfseries},
		cells={font=\footnotesize},
		}
		\toprule
		Medium & Role                                    & $\|\mathrm{std}(\epsilon_r)\|/\|\mathbb{E}\epsilon_r\|$ & $\|\mathrm{std}(k)\|/\|\mathbb{E}k\|$ & $\bar{\rho}_{\mathrm{ph}}$ \\
		\midrule
		S1     & Low-loss propagation-dominated baseline & 0.031                                                   & 0.016                                 & 0.357                      \\
		S2     & Strongly dispersive anchor              & 0.191                                                   & 0.092                                 & 4.260                      \\
		S3     & Ice/frozen-ground anchor                & 0.049                                                   & 0.025                                 & 0.581                      \\
		S4     & Lossy engineering anchor                & 0.055                                                   & 0.027                                 & 2.176                      \\
		S5     & Fine-grained lossy soil anchor          & 0.038                                                   & 0.019                                 & 1.389                      \\
		\bottomrule
	\end{tblr}
\end{table*}

\subsubsection{gprMax-Based Full-Wave Snapshot Setup}
\label{sec:gprmax_snapshot_setup}

An additional full-wave synthetic consistency check is conducted using \texttt{gprMax} to check whether the dominant covariance spectral signatures predicted by the proposed model can also be observed in frequency-domain snapshots generated by an independent FDTD electromagnetic solver. Relevant paramter settings are summarized in \Cref{tbl:gprmax_config}.

The compatibility layer constructs an FDA-MIMO-like snapshot from multistatic gprMax traces. For each realization, time-domain responses from a $4\times 4$ transmitter--receiver configuration are transformed to the frequency domain, and 25 DFT bins in the $200$--$300$ MHz band are retained. This gives a raw $4\times 4\times 25$ complex tensor. A fixed frequency-band compression is then applied to obtain an $M=16$ space--frequency vector, matching the covariance dimension used in the analytical validation.

\begin{table}[!t]
	\centering
	\caption{Configuration of the gprMax-based full-wave synthetic consistency check.}
	\label{tbl:gprmax_config}
	\begin{tblr}{
		width=\columnwidth,
		colspec={X[1.25,l]X[1.65,r]},
		row{1}={font=\bfseries},
		cells={font=\footnotesize},
		}
		\toprule
		Item                 & Setting                                            \\
		\midrule
		Solver               & \texttt{gprMax} v4.0 FDTD solver                   \\
		Domain               & $0.80\times0.50\times0.80~\mathrm{m}^3$            \\
		Grid spacing         & $0.02~\mathrm{m}$                                  \\
		Time window          & $40~\mathrm{ns}$                                   \\
		Boundary condition   & 8-layer CPML                                       \\
		Array layout         & $4\times4$ multistatic channels                    \\
		Frequency band       & $200$--$300~\mathrm{MHz}$                          \\
		Frequency samples    & 25 DFT bins                                        \\
		Raw snapshot tensor  & $4\times4\times25$                                 \\
		Compressed dimension & $M=16$                                             \\
		Dispersive material  & Cole--Cole approximated by 12-pole Debye expansion \\
		Level A perturbation & Cole--Cole parameter perturbation                  \\
		Level B perturbation & Debye-pole relaxation-spectrum perturbation        \\
		\bottomrule
	\end{tblr}
\end{table}

Two medium anchors, S1 and S5 from \Cref{tbl:medium_summary} are used. Two perturbation routes are tested for potential medium-prior mismatches. Level A perturbs the Cole--Cole parameter family directly and therefore represents a low-dimensional material-prior perturbation. Level B perturbs the Debye-pole relaxation spectrum after the Cole--Cole-to-Debye approximation and therefore produces a higher-dimensional spectral perturbation. The distinction is useful because the proposed theory is formulated at the relaxation-spectrum level, whereas many material priors are available as low-dimensional Cole--Cole parameters.

The gprMax snapshots are used differently from the model-based statistical simulations generated from \Cref{sec:analytical_simulation_setup}. \texttt{gprMax} internally solves the FDTD full-wave problem and provides final time-domain traces, from which frequency-domain FDA-MIMO-like snapshots are extracted by the compatibility layer. Hence, the full-wave data are used only at the snapshot and covariance levels. They mainly support the snapshot-level covariance spectral-structure comparison in \Cref{sec:monte_carlo_closure}, the full-wave Level-A CV perturbation response discussed alongside the validity-envelope analysis in \Cref{sec:validity_envelope}, the covariance-aware detection and whitening trend in \Cref{sec:detection_whitening}, and the prior-mismatch interpretation in \Cref{sec:mismatch_robustness}.

\subsection{Degeneration-Based Model-Consistency Checks}\label{sec:degeneration_checks}

Before quantitative testing, the model is checked against simpler limiting descriptions. The degeneration tests verify correct reduction when modeling ingredients are removed. These checks directly test the linear operator chain in Section~\ref{subsec:random_to_steering} and the covariance decomposition in Section~\ref{subsec:covariance_decomposition}.

If the model is a legitimate extension, it should reduce when randomness, dispersion, frequency diversity, or multi-channel structure is suppressed:
\begin{equation*}
	\delta \tilde g(u)\rightarrow 0,\qquad
	\epsilon_c(\omega)\rightarrow \epsilon_r,\qquad
	f_m\rightarrow f_0,\qquad
	M\rightarrow 1
	\label{eq:limiting_cases}
\end{equation*}

All four checks pass at machine precision (\Cref{tbl:limiting_cases}). The first two rows verify the intended removal of propagation-side randomness and dispersion, and the last two verify reduction to simpler array-structure limits.

\begin{table}[!t]
	\centering
	\caption{Degeneration-based model-consistency checks. Each case removes one modeling ingredient and verifies reduction to the corresponding simpler description.}
	\label{tbl:limiting_cases}
	\begin{tblr}{
		width=\columnwidth,
		colspec={X[1.3,l]X[1.4,l]X[0.8,c]X[1.8,l]},
		row{1}={font=\bfseries},
		cells={font=\footnotesize},
		}
		\toprule
		Case                  & Removed ingredient                       & Metric value & Expected reduction                         \\
		\midrule
		No medium uncertainty & Random medium perturbation               & $0$          & $\bm{R}_{\mathrm{med}} \rightarrow \bm{0}$ \\
		Non-dispersive medium & Frequency-dependent dispersion           & $0$          & $k=\omega\sqrt{\epsilon_r}/c$              \\
		Same-frequency MIMO   & FDA frequency coding                     & $0$          & FDA-specific coupling removed              \\
		Single-channel GPR    & Multi-channel space--frequency structure & $0$          & Scalar point-scattering response           \\
		\bottomrule
	\end{tblr}
\end{table}

\subsection{Monte Carlo Closure of the Statistical Propagation Chain}\label{sec:monte_carlo_closure}

The synthetic-prior baseline is used to verify the internal closure of the operator chain from relaxation-spectrum perturbation to total clutter covariance. Following the theoretical chain in Section~\ref{sec:theory} (Assumptions~\ref{asm:second_order} through~\ref{asm:first_order_phase}), the closure is checked stage by stage:
\begin{equation*}
	\begin{gathered}
		\delta\tilde g(u)\rightarrow \delta\epsilon_c(\omega)\rightarrow \delta k_c(\omega)\rightarrow \delta \bm a(\theta,r)\\
		\rightarrow \bm R_a(\theta,r)\rightarrow \bm R_{\mathrm{med}}\rightarrow \bm R_c
		\rightarrow r_{\mathrm{eff}},\,p_\rho,\,\eta_\rho,\,\gamma_\rho
	\end{gathered}
	\label{eq:closure_chain}
\end{equation*}
The objective is to identify where the dominant approximation error enters the chain. Results are summarized in \Cref{tbl:baseline_closure_metrics,tbl:baseline_spectral_closure}. At the front end, $e_{\epsilon}=1.49\times10^{-2}$ and $e_k=1.49\times10^{-2}$. At the steering stage, $e_{a,\max}=1.41\times10^{-2}$ and the maximum RMS steering-linearization error is $7.11\times10^{-2}$. At the scene level, $e_{\mathrm{prop}}=1.39\times10^{-2}$ and $e_c=4.58\times10^{-3}$. No numerical instability is observed: the branch-flip count is zero, the Hermitian residual is $6.12\times10^{-18}$, and the minimum eigenvalue remains at machine precision.

\begin{table*}[!t]
	\centering
	\caption{Baseline stage-wise Monte Carlo closure metrics for the main statistical propagation chain.}
	\label{tbl:baseline_closure_metrics}
	\begin{tblr}{
		width=\textwidth,
		colspec={X[0.6,c]X[2.2,l]X[0.9,c]X[0.9,c]},
		row{1}={font=\bfseries},
		cells={font=\footnotesize},
		}
		\toprule
		Stage & Metric                                                               & Value                 & Acceptance level       \\
		\midrule
		A     & Relative error of permittivity covariance $e_{\epsilon}$             & $1.49\times10^{-2}$   & $<5\times10^{-2}$      \\
		B     & Relative error of wavenumber covariance $e_k$                        & $1.49\times10^{-2}$   & $<5\times10^{-2}$      \\
		B     & RMS relative error of first-order wavenumber linearization           & $4.30\times10^{-3}$   & $<5\times10^{-2}$      \\
		C     & Maximum relative error of local steering covariance $e_{a,\max}$     & $1.41\times10^{-2}$   & $<8\times10^{-2}$      \\
		C     & Maximum RMS relative error of steering linearization                 & $7.11\times10^{-2}$   & $<8\times10^{-2}$      \\
		D     & Relative error of propagation-induced covariance $e_{\mathrm{prop}}$ & $1.39\times10^{-2}$   & $<8\times10^{-2}$      \\
		D     & Relative error of total clutter covariance $e_c$                     & $4.58\times10^{-3}$   & $<8\times10^{-2}$      \\
		D     & Maximum Hermitian residual                                           & $6.12\times10^{-18}$  & $<10^{-12}$            \\
		D     & Minimum eigenvalue                                                   & $-3.84\times10^{-18}$ & near machine precision \\
		D     & Branch-flip count in $k_c$ mapping                                   & 0                     & no branch switching    \\
		\bottomrule
	\end{tblr}
\end{table*}

Eigendecomposition confirms that the covariance-level agreement carries through to the spectral diagnostics in Section~\ref{subsec:spectral_broadening_rank} and Section~\ref{subsec:subspace_separability}: $r_{\mathrm{eff}}$ differs by $1.4\times10^{-2}$, both routes yield $p_\rho=3$ at $\rho=0.9$, and the separability measure changes from $0.654$ to $0.651$.

\begin{table}[!t]
	\centering
	\caption{Baseline spectral and separability consistency derived from $\bm R_c$.}
	\label{tbl:baseline_spectral_closure}
	\begin{tblr}{
		width=\columnwidth,
		colspec={X[1.8,l]X[0.9,c]X[0.9,c]},
		row{1}={font=\bfseries},
		cells={font=\footnotesize},
		}
		\toprule
		Metric                                              & Theory & Monte Carlo \\
		\midrule
		Effective rank $r_{\mathrm{eff}}$                   & 2.761  & 2.775       \\
		Effective subspace dimension $p_\rho$ at $\rho=0.9$ & 3      & 3           \\
		Representative separability $\eta_\rho$             & 0.654  & 0.651       \\
		Representative complement $\gamma_\rho$             & 0.346  & 0.349       \\
		Subspace instability flag                           & false  & false       \\
		\bottomrule
	\end{tblr}
\end{table}

Scanning $N_{\mathrm{MC}}$ over $\{200,500,1000,2000\}$ (\cref{tbl:nmc_closure_convergence}) shows that the front-end errors decrease from about $1.2\times10^{-1}$ at $N_{\mathrm{MC}}=200$ to about $1.5\times10^{-2}$ at $N_{\mathrm{MC}}=2000$, while $e_c$ remains near $4.6\times10^{-3}$. In the present setup, $N_{\mathrm{MC}}=1000$ already yields few-percent closure errors, and $2000$ provides additional margin.

\begin{table*}[!t]
	\centering
	\caption{Monte Carlo convergence of representative closure metrics with fixed $\sigma_g=0.03$.}
	\label{tbl:nmc_closure_convergence}
	\begin{tblr}{
		width=\textwidth,
		colspec={X[0.8,c]X[1.0,c]X[1.0,c]X[1.0,c]X[1.0,c]X[1.0,c]},
		row{1}={font=\bfseries},
		cells={font=\footnotesize},
		}
		\toprule
		$N_{\mathrm{MC}}$ & $e_{\epsilon}$ & $e_k$  & $e_{a,\max}$ & $e_{\mathrm{prop}}$ & $e_c$   \\
		\midrule
		200               & 0.1226         & 0.1204 & 0.0973       & 0.1046              & 0.00454 \\
		500               & 0.0518         & 0.0505 & 0.0350       & 0.0399              & 0.00456 \\
		1000              & 0.0370         & 0.0370 & 0.0346       & 0.0352              & 0.00456 \\
		2000              & 0.0149         & 0.0149 & 0.0141       & 0.0139              & 0.00458 \\
		\bottomrule
	\end{tblr}
\end{table*}

In addition to the analytical Monte Carlo closure, the gprMax-based full-wave snapshots described in \Cref{sec:gprmax_snapshot_setup} are used to check whether the same final covariance spectral signature is observed at the snapshot level.

\begin{table}[!t]
	\centering
	\caption{Spectral consistency between the analytical covariance, analytical Monte Carlo closure, and gprMax-based full-wave synthetic snapshots.}
	\label{tbl:gprmax_spectral_consistency}
	\begin{tblr}{
		width=\columnwidth,
		colspec={X[1.6,l]X[0.7,c]X[0.7,c]X[0.8,c]X[0.8,c]},
		row{1}={font=\bfseries},
		cells={font=\footnotesize},
		}
		\toprule
		Metric                                     & Theory & MC    & FW-A  & FW-B  \\
		\midrule
		Effective rank $r_{\mathrm{eff}}$          & 2.761  & 2.775 & 2.810 & 3.139 \\
		Effective dimension $p_\rho$ at $\rho=0.9$ & 3      & 3     & 3     & 3     \\
		Subspace instability flag                  & false  & false & false & false \\
		\bottomrule
	\end{tblr}
\end{table}

As shown in \Cref{tbl:gprmax_spectral_consistency}, Level A full-wave snapshots give $r_{\mathrm{eff}}=2.810$, close to the analytical value $2.761$ and the analytical Monte Carlo value $2.775$. Level B gives a larger effective rank, $r_{\mathrm{eff}}=3.139$, which is consistent with its higher-dimensional relaxation-spectrum perturbation. In both full-wave routes, the effective dimension remains $p_\rho=3$, matching the analytical prediction. This agreement supports the dominant covariance spectral signature at the final snapshot level.

\subsection{First-Order Validity Envelope}\label{sec:validity_envelope}

To specify how far the first-order approximation can be pushed under Assumption~\ref{asm:first_order_phase}, the perturbation strength is scaled from $0.25\times$ to $3.0\times$ baseline while monitoring $e_{\mathrm{lin}}$, $e_{R_c}$, and $e_r$.

A phase-perturbation proxy $\tilde\rho = \kappa\sigma_g\bar{L}$ serves as the organizing variable, where $\kappa$ is the perturbation scale factor, $\sigma_g$ is the baseline spectral perturbation amplitude, and $\bar{L}$ is the domain-averaged propagation path length. This proxy is a monotonic scalar surrogate for the rigorous $\rho_{\mathrm{ph}}$ defined in \eqref{eq:rho_ph_def_assumption}, which would require full integration over the wavenumber-perturbation covariance. It is not directly comparable to the per-medium $\bar{\rho}_{\mathrm{ph}}$ reported in \cref{tbl:medium_summary}, which originates from Cole--Cole parameter sampling rather than the Mat\'ern-based $\delta\tilde g$ model.

The trend in \Cref{tbl:validity_envelope} is monotone. The first four points ($\rho_{\mathrm{ph}}\le 0.0826$) remain in a weak regime with $e_{\mathrm{lin}}\le 7.25\times10^{-2}$ and $e_{R_c}\le 4.62\times10^{-3}$. The next three enter a moderate regime: at scale $3.0$, $e_{\mathrm{lin}}$ reaches $2.11\times10^{-1}$ while $e_{R_c}=4.00\times10^{-2}$ and $e_r=4.50\times10^{-2}$. Across the full scan, the local linearization error grows faster than the aggregated covariance and rank errors.

\begin{table*}[!t]
	\centering
	\caption{First-order validity envelope under perturbation scaling. The $\rho_{\mathrm{ph}}$ proxy $\tilde\rho = \kappa\sigma_g\bar{L}$ is a monotonic scalar surrogate for the rigorous $\rho_{\mathrm{ph}}$ in \eqref{eq:rho_ph_def_assumption} and differs from the per-medium $\bar{\rho}_{\mathrm{ph}}$ in \protect\cref{tbl:medium_summary}, which originates from Cole--Cole parameter sampling.}
	\label{tbl:validity_envelope}
	\begin{tblr}{
		width=\textwidth,
		colspec={X[0.9,c]X[1.1,c]X[1.0,c]X[1.0,c]X[0.9,c]X[1.0,c]},
		row{1}={font=\bfseries},
		cells={font=\footnotesize},
		}
		\toprule
		Scale & $\rho_{\mathrm{ph}}$ proxy & $e_{\mathrm{lin}}$ & $e_{R_c}$            & $e_r$                & Label    \\
		\midrule
		0.25  & 0.0206                     & 0.0182             & $2.89\times 10^{-4}$ & $3.36\times 10^{-4}$ & weak     \\
		0.50  & 0.0413                     & 0.0364             & $1.16\times 10^{-3}$ & $1.34\times 10^{-3}$ & weak     \\
		0.75  & 0.0619                     & 0.0545             & $2.60\times 10^{-3}$ & $3.02\times 10^{-3}$ & weak     \\
		1.00  & 0.0826                     & 0.0725             & $4.62\times 10^{-3}$ & $5.36\times 10^{-3}$ & weak     \\
		1.50  & 0.1239                     & 0.1082             & $1.03\times 10^{-2}$ & $1.20\times 10^{-2}$ & moderate \\
		2.00  & 0.1652                     & 0.1432             & $1.82\times 10^{-2}$ & $2.10\times 10^{-2}$ & moderate \\
		3.00  & 0.2478                     & 0.2111             & $4.00\times 10^{-2}$ & $4.50\times 10^{-2}$ & moderate \\
		\bottomrule
	\end{tblr}
\end{table*}

The gprMax snapshots cannot provide the intermediate perturbation terms needed for $e_{\mathrm{lin}}$ or $e_{R_c}$, but they can show the final covariance-spectrum response to material perturbations. \Cref{tbl:gprmax_levelA_cv_scan} reports a Level-A Cole--Cole scan, where CV is the coefficient of variation, i.e., the relative standard deviation used to perturb the Cole--Cole material parameters.

\begin{table}[!t]
	\centering
	\caption{Level-A full-wave CV scan using compressed 16-D gprMax covariance snapshots. CV denotes the coefficient of variation of the Cole--Cole parameter perturbations.}
	\label{tbl:gprmax_levelA_cv_scan}
	\begin{tblr}{
		width=\columnwidth,
		colspec={X[0.65,c]X[0.7,c]X[0.65,c]X[0.9,c]X[0.9,c]X[0.95,c]},
		row{1}={font=\bfseries},
		cells={font=\footnotesize},
		}
		\toprule
		Medium & CV   & $N$ & $r_{\mathrm{eff}}$ & $p_\rho$ & $\lambda_1/\sum_i\lambda_i$ \\
		\midrule
		S1     & 0.03 & 3   & 1.474              & 2        & 0.799                       \\
		S1     & 0.05 & 5   & 1.857              & 2        & 0.680                       \\
		S1     & 0.10 & 5   & 2.019              & 2        & 0.642                       \\
		S1     & 0.20 & 5   & 2.252              & 2        & 0.564                       \\
		S1     & 0.30 & 5   & 1.498              & 2        & 0.804                       \\
		\midrule
		S5     & 0.03 & 5   & 1.007              & 1        & 0.996                       \\
		S5     & 0.05 & 4   & 1.170              & 1        & 0.921                       \\
		S5     & 0.10 & 5   & 1.005              & 1        & 0.997                       \\
		S5     & 0.20 & 5   & 1.012              & 1        & 0.994                       \\
		S5     & 0.30 & 5   & 1.006              & 1        & 0.997                       \\
		\bottomrule
	\end{tblr}
\end{table}

The full-wave scan shows a medium-dependent covariance-spectrum response. For the low-loss S1 anchor, increasing the Level-A CV from $0.03$ to $0.20$ broadens the compressed covariance spectrum: $r_{\mathrm{eff}}$ increases from $1.474$ to $2.252$, and the first-eigenvalue energy fraction decreases from $0.799$ to $0.564$. The CV $=0.30$ case is nonmonotone, returning to a more concentrated spectrum with $r_{\mathrm{eff}}=1.498$ and $\lambda_1/\sum_i\lambda_i=0.804$.

For the lossy S5 anchor, the compressed covariance remains nearly rank one for all tested CV values, with $p_\rho=1$ and $r_{\mathrm{eff}}$ close to unity. The first eigenvalue contains more than $0.92$ of the covariance energy in all cases. This contrast suggests that low-loss media preserve richer space--frequency perturbation modes, whereas lossy media compress observable full-wave variability into a dominant covariance mode.

\subsection{Covariance-Aware Detection and Whitening}\label{sec:detection_whitening}

This section tests whether the induced covariance benefits downstream detection in a covariance-aware weak-target problem, following the separability framework developed in Section~\ref{subsec:subspace_separability}. The task is posed under the binary hypotheses
\begin{equation}
	\mathcal H_0:\ \bm y=\bm c+\bm n,\qquad
	\mathcal H_1:\ \bm y=\alpha_t \bm a_t+\bm c+\bm n
	\label{eq:hypotheses}
\end{equation}
where $\bm a_t$ is the nominal steering vector of the representative target patch, $\alpha_t$ is the target amplitude, $\bm c$ is the clutter term with covariance $\bm R_c$, and $\bm n$ is additive noise. In the current benchmark, the Monte Carlo clutter covariance $\bm R_c^{\mathrm{mc}}$ is treated as the reference clutter covariance, and the noise level is set by
\begin{equation}
	\sigma_n^2
	=
	10^{-3}\,
	\frac{\mathrm{tr}(\bm R_c^{\mathrm{mc}})}{M}
	\label{eq:noise_variance}
\end{equation}
The target amplitude is then chosen so that the input SCR equals $-10$ dB:
\begin{equation}
	|\alpha_t|^2
	=
	10^{-10/10}\,
	\frac{\mathrm{tr}(\bm R_c^{\mathrm{mc}})}
	{\bm a_t^{\mathrm H}\bm a_t}
	\label{eq:target_amplitude}
\end{equation}

Operationally, the processor is supplied with a candidate covariance model $\widehat{\bm R}$ and forms the whitened observation
\begin{equation}
	\widetilde{\bm y}
	=
	\widehat{\bm R}^{-1/2}\bm y
	\label{eq:whitened_obs}
\end{equation}
equivalent to a covariance-aware matched filter with weight
\begin{equation}
	\bm w
	=
	\widehat{\bm R}^{-1}\bm a_t
	\label{eq:matched_filter}
\end{equation}
The resulting detection statistic is
\begin{equation}
	T(\bm y;\widehat{\bm R})
	=
	\frac{
	\left|
	\bm a_t^{\mathrm H}\widehat{\bm R}^{-1}\bm y
	\right|^2
	}{
	\bm a_t^{\mathrm H}\widehat{\bm R}^{-1}\bm a_t
	}
	\label{eq:detection_statistic}
\end{equation}
evaluated separately on target-absent and target-present Monte Carlo draws. This setup serves only as a controlled task-level probe for testing whether the modeled cross-channel covariance structure is practically informative once inserted into a standard whitening-and-detection pipeline.

The benchmark compares seven processor-side covariance choices: the identity matrix, the nominal background $\bm R_0$, a purely diagonal approximation, a block-diagonal approximation, the proposed full covariance, an oracle covariance, and the propagation-induced term alone. The metrics span three aspects of performance: AUC for ranking quality, fixed-false-alarm detection probability ($P_D@10^{-2}$) for operating-point behavior, and whitening error $e_{\mathrm{white}}$ for covariance fidelity, through
\begin{equation}
	e_{\mathrm{white}}
	=
	\frac{
	\left\|
	\widehat{\bm R}^{-1/2}\bm R_{\mathrm{true}}\widehat{\bm R}^{-{\mathrm H}/2}
	-\bm I
	\right\|_{\mathrm F}
	}{
	\|\bm I\|_{\mathrm F}
	}
	\label{eq:whitening_error}
\end{equation}
where $\bm R_{\mathrm{true}}=\bm R_c^{\mathrm{mc}}+\sigma_n^2\bm I$. Together these metrics distinguish whether a covariance helps detection, helps whitening, or merely captures nominal background structure.

As shown in \Cref{tbl:detection_summary}, the structured covariance models all achieve AUC values of $0.753$--$0.754$ and detection probabilities $P_D@10^{-2}$ of $0.126$--$0.134$, whereas the identity and diagonal baselines reach AUC values of $0.613$ and $0.593$, respectively. For whitening fidelity, the non-oracle ranking by $e_{\mathrm{white}}$ is block-diagonal ($1.722$), diagonal ($2.140$), proposed ($2.346$), and $\bm R_0$ ($2.356$); the identity model yields $2401.345$.

\begin{table*}[!t]
	\centering
	\caption{Covariance-aware detection and whitening summary for the representative benchmark.}
	\label{tbl:detection_summary}
	\begin{tblr}{
		width=\textwidth,
		colspec={X[1.5,l]X[0.9,c]X[1.0,c]X[1.0,c]X[1.0,c]},
		row{1}={font=\bfseries},
		cells={font=\footnotesize},
		}
		\toprule
		Model          & AUC   & $P_D@10^{-2}$ & $e_{\mathrm{white}}$ & $\Delta\mathrm{AUC}$ vs. diagonal \\
		\midrule
		Identity       & 0.613 & 0.040         & 2401.345             & 0.020                             \\
		$\bm R_0$      & 0.753 & 0.127         & 2.356                & 0.160                             \\
		Diagonal       & 0.593 & 0.029         & 2.140                & 0.000                             \\
		Block-diagonal & 0.754 & 0.134         & 1.722                & 0.162                             \\
		Proposed       & 0.753 & 0.126         & 2.346                & 0.160                             \\
		Oracle         & 0.775 & 0.153         & $3.0\times 10^{-13}$ & 0.182                             \\
		\bottomrule
	\end{tblr}
\end{table*}

The gprMax snapshots described in \Cref{sec:gprmax_snapshot_setup} are used as a full-wave trend probe for the same whitening-and-detection pipeline. With only a limited number of FDTD realizations, the full covariance estimate is shrinkage-regularized. Because structured covariance processors yield saturated detection scores in this small-sample setting, the probe is used to check trend consistency.

\begin{table*}[!t]
	\centering
	\caption{gprMax-based full-wave detection and whitening trend. The table reports ranges over Level A/B perturbations and scatter/background covariance probes. Saturated means that all structured covariance processors reach the numerical upper bound in this small-sample full-wave probe.}
	\label{tbl:gprmax_detection_probe}
	\begin{tblr}{
		width=\textwidth,
		colspec={X[1.5,l]X[1.0,c]X[1.0,c]X[1.1,c]X[1.2,c]},
		row{1}={font=\bfseries},
		cells={font=\footnotesize},
		}
		\toprule
		Model          & AUC range    & $P_D@10^{-2}$ & $e_{\mathrm{white}}$ range & Interpretation                                \\
		\midrule
		Identity       & 0.522--0.698 & low           & 2.022--3.811               & weak unstructured baseline                    \\
		Diagonal       & saturated    & saturated     & 0.781--2.321               & covariance-aware but channel-decoupled        \\
		Block-diagonal & saturated    & saturated     & 0.678--0.889               & best non-oracle structured whitening          \\
		Proposed       & saturated    & saturated     & 0.757--0.864               & full covariance with shrinkage regularization \\
		Oracle         & saturated    & saturated     & 0.674--0.772               & oracle-shrinkage reference                    \\
		\bottomrule
	\end{tblr}
\end{table*}

As shown in \Cref{tbl:gprmax_detection_probe}, the identity model is the weakest processor for the full-wave snapshots, with AUC values of $0.522$--$0.698$ and substantially larger whitening errors. In contrast, the covariance-aware models reach saturated detection scores in this small-sample probe. The whitening metric still provides coarse separation among structured models: the block-diagonal, proposed full-covariance, and oracle models all reduce $e_{\mathrm{white}}$ well below the identity baseline.

These results are interpreted only as full-wave trend-level support. They indicate that the gprMax snapshots contain exploitable covariance structure for whitening and detection, but they do not provide a reliable ranking among structured covariance processors. The controlled analytical benchmark in \Cref{tbl:detection_summary} remains the main comparison for processor-level performance.

\subsection{Robustness to Medium-Prior Mismatch}\label{sec:mismatch_robustness}

An application-level covariance model is useful only if its gains persist under modest prior misspecification. The final task-level experiment therefore studies processor mismatch, testing the propagation-dominated approximation discussed in Section~\ref{subsec:propagation_dominated_scope}: the clutter-generating model is fixed, while the processor is supplied with perturbed medium-prior parameters. Concretely, the experiment replaces the matched prior $(\sigma_g,\ell)$ by a processor-side pair
\begin{equation}
	\widehat{\sigma}_g=\kappa_\sigma \sigma_g,\qquad
	\widehat{\ell}=\kappa_\ell \ell
	\label{eq:mismatch_prior}
\end{equation}
with $(\kappa_\sigma,\kappa_\ell)$ scanned over a two-dimensional grid. For each pair, the processor reconstructs a new covariance
\begin{equation}
	\widehat{\bm R}_{c}^{\mathrm{prop}}
	(\kappa_\sigma,\kappa_\ell)
	\label{eq:mismatch_covariance}
\end{equation}
from the same first-order propagation chain, and then re-runs the whitening-and-detection pipeline based on the same statistic
\begin{equation}
	T\bigl(\bm y;\widehat{\bm R}_{c}^{\mathrm{prop}}(\kappa_\sigma,\kappa_\ell)\bigr)
	\label{eq:mismatch_statistic}
\end{equation}
This tests whether the proposed method behaves like a brittle tuned model or a controllably biased prior-informed processor.

Besides the task-level metrics, the mismatch experiment also records a direct covariance error,
\begin{equation}
	e_{\mathrm{mis}}
	=
	\frac{
	\left\|
	\widehat{\bm R}_{c}^{\mathrm{prop}}(\kappa_\sigma,\kappa_\ell)
	-\bm R_c^{\mathrm{mc}}
	\right\|_{\mathrm F}
	}{
	\left\|
	\bm R_c^{\mathrm{mc}}
	\right\|_{\mathrm F}
	}
	\label{eq:mismatch_error}
\end{equation}
so that changes in AUC can be interpreted together with the amount of processor-side covariance misspecification.

\begin{table*}[!t]
	\centering
	\caption{Selected prior-mismatch results: proposed AUC, gain over the diagonal baseline, and covariance mismatch error.}
	\label{tbl:mismatch_summary}
	\begin{tblr}{
		width=\textwidth,
		colspec={X[1.5,l]X[0.7,c]X[0.7,c]X[0.9,c]X[1.2,c]X[1.2,c]},
		row{1}={font=\bfseries},
		cells={font=\footnotesize},
		}
		\toprule
		Scenario       & $\kappa_\sigma$ & $\kappa_\ell$ & AUC   & $\Delta$AUC vs.\ diag. & $e_{\mathrm{mis}}$ \\
		\midrule
		Matched        & 1.00            & 1.00          & 0.738 & 0.168                  & 0.0063             \\
		Mild over      & 1.25            & 1.25          & 0.754 & 0.163                  & 0.0093             \\
		Moderate over  & 1.50            & 1.50          & 0.759 & 0.179                  & 0.0142             \\
		Severe over    & 2.00            & 2.00          & 0.736 & 0.155                  & 0.0279             \\
		Moderate under & 0.50            & 0.50          & 0.749 & 0.183                  & 0.0059             \\
		Severe under   & 0.25            & 0.25          & 0.743 & 0.151                  & 0.0063             \\
		Worst-case AUC & 1.50            & 0.50          & 0.733 & 0.147                  & 0.0091             \\
		\bottomrule
	\end{tblr}
\end{table*}

\begin{figure}[!t]
	\centering
	\includegraphics[width=\columnwidth]{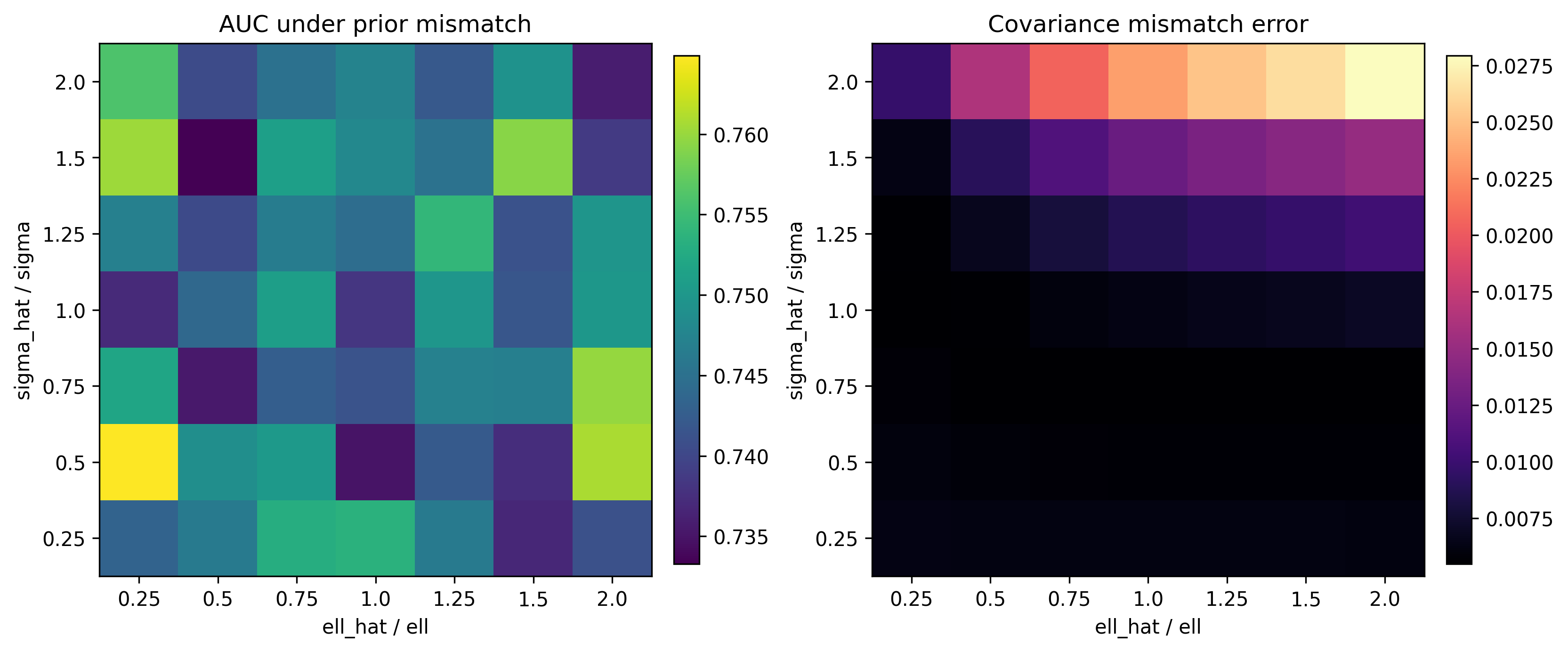}
	\caption{Prior-mismatch robustness: proposed AUC (left) and covariance mismatch error (right) across the mismatch grid. Mild mismatch preserves gain over diagonal; severe mismatch causes gradual loss.}
	\label{fig:exp_mismatch_heatmap}
\end{figure}

Two regimes are visible in \cref{tbl:mismatch_summary,fig:exp_mismatch_heatmap}. Under mild mismatch ($\kappa_\sigma,\kappa_\ell$ within $0.75$--$1.25$), the proposed processor preserves a mean AUC gain of $0.1605$ over the diagonal baseline, with individual AUC values of $0.738$--$0.754$ across the nine points in this window. Under severe mismatch---ratios $\ge 1.5$ or $\le 0.5$ for at least one parameter---the mean AUC decreases to $0.746$, and at the worst-case grid point ($\kappa_\sigma=1.5$, $\kappa_\ell=0.5$) the AUC is $0.733$ with $\Delta$AUC $=0.147$ over the diagonal baseline.

The covariance mismatch error $e_{\mathrm{mis}}$ is also asymmetric between under- and overestimation. For $\kappa_\sigma,\kappa_\ell<1$, $e_{\mathrm{mis}}$ remains within $0.0055$--$0.0063$; for $\kappa_\sigma,\kappa_\ell>1$, it increases from $0.0063$ at the matched point to $0.0279$ at full overestimation. The heatmap further shows stronger AUC sensitivity to $\ell$ mismatch than to $\sigma_g$ mismatch.

The gprMax snapshots provide a complementary full-wave mismatch check at the covariance-subspace level. Unlike the analytical grid in \Cref{tbl:mismatch_summary}, which perturbs $(\sigma_g,\ell)$ within the same statistical propagation model, the full-wave check compares processor covariances estimated from different material-prior families. The generator is fixed to S1 CV $=0.05$, and the processor covariance is replaced by either another S1 prior or an S5 prior.

\begin{table*}[!t]
	\centering
	\caption{gprMax-based full-wave prior-mismatch behavior. The generator covariance is estimated from S1 CV $=0.05$ snapshots, and each processor covariance is estimated from the listed medium prior.}
	\label{tbl:gprmax_mismatch_summary}
	\begin{tblr}{
		width=\textwidth,
		colspec={X[1.25,l]X[0.65,c]X[0.8,c]X[0.8,c]X[0.7,c]X[0.75,c]X[1.15,c]X[0.8,c]},
		row{1}={font=\bfseries},
		cells={font=\footnotesize},
		}
		\toprule
		Processor prior
		              & $N$
		              & $e_{\mathrm{mis}}$
		              & $r_{\mathrm{eff}}$
		              & Generator $p_\rho$
		              & Processor $p_\rho$
		              & Principal angles $(^\circ)$
		              & Subspace overlap                                                                  \\
		\midrule
		S1 CV $=0.10$ & 5                           & 1.124    & 2.207 & 3 & 3 & 1.1, 3.5, 4.8    & 0.985 \\
		S1 CV $=0.30$ & 5                           & 1.421    & 1.643 & 3 & 3 & 1.4, 3.4, 14.6   & 0.925 \\
		S5 CV $=0.10$ & 5                           & 566.982  & 1.106 & 3 & 1 & 15.4, 33.6, 57.4 & 0.911 \\
		S5 CV $=0.30$ & 5                           & 1567.784 & 1.108 & 3 & 1 & 21.0, 27.4, 50.9 & 0.688 \\
		\bottomrule
	\end{tblr}
\end{table*}

The full-wave mismatch behavior in \Cref{tbl:gprmax_mismatch_summary} distinguishes within-family and cross-family prior errors. When the S1 generator is processed with another S1 prior, $e_{\mathrm{mis}}$ remains in the range $1.124$--$1.421$, the effective dimension is preserved at $p_\rho=3$, and the subspace overlap remains high at $0.925$--$0.985$. The first two principal angles are also small. Thus, changing the perturbation strength within the same low-loss material family alters the covariance energy distribution but does not destroy the dominant full-wave covariance subspace.

The S5 processor priors show a stronger mismatch effect. The mismatch error increases to $566.982$ and $1567.784$, and the processor effective dimension collapses from the generator value $p_\rho=3$ to $p_\rho=1$. The principal angles also increase, and the subspace overlap decreases, especially for S5 CV $=0.30$. This result is consistent with the analytical mismatch scan: the covariance-aware processor is robust to moderate prior errors but not to arbitrary material-class mismatch. Because the full-wave covariance estimates and analytical mismatch grid use different data-generation mechanisms, the absolute values of $e_{\mathrm{mis}}$ are not compared across the two tables.

\subsection{Discussion}\label{sec:discussion}

The experiments support a coherent but bounded reading of the theory in \cref{sec:theory}. The setup in \cref{sec:numerical_setup,tbl:numerical_config,tbl:medium_parameters,tbl:medium_summary} anchors the numerical study to a physically traceable prior family. The degeneration checks in \cref{sec:degeneration_checks,tbl:limiting_cases} confirm correct reduction when randomness, dispersion, FDA frequency coding, or multi-channel structure is suppressed, as expected from \cref{subsec:random_to_steering,subsec:covariance_decomposition}. The closure results in \cref{sec:monte_carlo_closure,tbl:baseline_closure_metrics,tbl:baseline_spectral_closure,tbl:nmc_closure_convergence} show that the propagation chain remains numerically consistent from $\delta\tilde g$ to $\bm R_c$, carrying through to the spectral and subspace quantities in \cref{subsec:spectral_broadening_rank,subsec:subspace_separability}. gprMax-based full-wave snapshots (\Cref{sec:gprmax_snapshot_setup}) corroborate this closure at the snapshot level (\Cref{tbl:gprmax_spectral_consistency}): Level-A snapshots yield $r_{\mathrm{eff}}=2.810$, close to the analytical $2.761$, and both Level A and B preserve $p_\rho=3$. This independent FDTD check confirms that the dominant spectral signature is not an artefact of the analytical chain. The experimental section thus supports the paper's central claim: the proposed model behaves as a controlled propagation-side extension of the nominal clutter formulation.

The clearest boundary of that claim is first-order steering linearization. In \cref{tbl:baseline_closure_metrics}, the largest stage-wise discrepancy is at Stage C, where the steering-linearization error exceeds the front-end covariance and assembled clutter-covariance errors. The same ordering appears in \cref{tbl:validity_envelope}, where $e_{\mathrm{lin}}$ grows faster than $e_{R_c}$ and $e_r$ as the perturbation scale increases. This hierarchy follows from \cref{eq:delta_a_m,eq:delta_a_vec,eq:R_c_decomp}: the mappings $\delta\tilde g\to\delta\epsilon_c$ and $\delta\epsilon_c\to\delta k_c$ are linear, whereas $\delta k_c$ affects the steering response through the exponential term in \cref{eq:a_m_expand}. Under Assumption~\ref{asm:first_order_phase}, the local steering perturbation is where nonlinearity first becomes visible. Thus, the weak regime in \cref{tbl:validity_envelope} is the most defensible quantitative range; the moderate regime remains informative for covariance and spectral summaries even as the local linearization degrades. The gprMax Level-A CV scan (\Cref{tbl:gprmax_levelA_cv_scan}) provides complementary full-wave evidence: for the low-loss S1 anchor, increasing CV from $0.03$ to $0.20$ broadens the compressed covariance spectrum ($r_{\mathrm{eff}}$ from $1.474$ to $2.252$), while the lossy S5 anchor stays near rank-one ($p_\rho=1$) across all CV values. This medium-dependent contrast aligns with the propagation-side hierarchy: low-loss media support longer paths and richer dispersion-induced modes, whereas lossy media compress variability into a single dominant mode.

The detection results clarify a second point: structured covariance matters, but the evidence does not yet establish that the full proposed covariance is uniquely necessary. In \cref{tbl:detection_summary}, the proposed model, $\bm R_0$, and the block-diagonal approximation are nearly indistinguishable in AUC and $P_D@10^{-2}$, despite different whitening errors. The gprMax detection probe (\Cref{tbl:gprmax_detection_probe}) is consistent: the identity model is the weakest processor across all full-wave configurations, while covariance-aware processors saturate. Although the probe does not reliably rank structured models, it confirms that exploitable covariance structure is present in FDTD-generated snapshots. Taken with the compact subspace dimension in \cref{tbl:baseline_spectral_closure} and \cref{eq:p_rho,eq:gamma_p,eq:eta_p}, this pattern suggests practical improvement is driven first by identifying the correct low-dimensional clutter structure and only second by refining within-subspace geometry. This aligns with the modal representation in \cref{eq:R_med_KL,eq:S_q}: if medium-induced structure concentrates in a few coupling patterns, a block-diagonal or nominal-background approximation captures most of the detection benefit despite not reproducing the full covariance detail.

This explains why the proposed covariance remains important despite modest raw detection gain over $\bm R_0$. The full model's contribution is not merely raising AUC in \cref{tbl:detection_summary}; it supplies an explicit mechanism linking dispersive-medium uncertainty to covariance broadening, rank growth, and subspace expansion through \cref{eq:R_c_decomp,eq:r_eff_total_mu,eq:prho_lower_bound}. The detection experiment thus tests whether the predicted covariance structure is operationally meaningful in a standard whitening-and-matched-filter pipeline (\cref{eq:hypotheses,eq:matched_filter,eq:detection_statistic}). The near-parity with simpler structured alternatives localizes where the practical value resides: the evidence supports the importance of medium-induced covariance structure, but suggests only part of it is needed for downstream processing in this benchmark.

The mismatch results (\cref{tbl:mismatch_summary,fig:exp_mismatch_heatmap}) sharpen the practical interpretation. The method behaves as a controlled prior-informed processor rather than a brittle tuned model: moderate mismatch preserves most of the gain, whereas severe mismatch causes gradual, not catastrophic, degradation. The asymmetry between under- and overestimation is theoretically meaningful: through \cref{eq:a_m_expand}, overestimation perturbs the steering response more aggressively, penalizing optimism more than conservatism. Nevertheless, the method remains prior-dependent; the covariance is only as credible as the medium parameters. The most plausible deployment thus requires reconnaissance-level dielectric characterization or site-specific prior knowledge. The gprMax mismatch check (\Cref{tbl:gprmax_mismatch_summary}) sharpens this by distinguishing within-family and cross-family prior errors. With an S1 generator and another S1 prior, $p_\rho=3$ and subspace overlap ($0.925$--$0.985$) remain high, confirming robustness within the same material family. Substituting an S5 prior collapses the effective dimension to $p_\rho=1$ and markedly reduces overlap. This reinforces the analytical finding: the processor is robust to in-family errors but not to arbitrary material-class mismatch, necessitating at least reconnaissance-level material-family identification.

The evidentiary scope of this section spans two tiers: the analytical Monte Carlo simulations (\Cref{sec:analytical_simulation_setup}) and the gprMax-based full-wave FDTD snapshots (\Cref{sec:gprmax_snapshot_setup}). The analytical tier provides fine-grained access to every intermediate propagation term, enabling quantitative closure and validity-envelope tests. The full-wave tier, though limited in snapshot count and constrained by compatibility-layer projection, supplies an independent solver-level check that confirms the analytical results at the snapshot and covariance levels. No measured FDA-MIMO-GPR data is used, and the detection benchmark is a controlled probe, not a mission-complete evaluation. This section establishes a reproducible validation envelope for the propagation-side covariance theory, supported by both analytical and full-wave evidence. This distinction matters because the two main theoretical boundaries remain active: the weak-phase requirement (Assumption~\ref{asm:first_order_phase}) and the propagation-dominated approximation (\cref{subsec:propagation_dominated_scope,eq:Rc_general_beta_random}). The natural next step is to stress both boundaries under richer scattering, stronger heterogeneity, and measured datasets---where the full-wave evidence already bridges part of the gap to field data, but where the omitted scattering term $\bm R_{\beta}$ and cross term $\bm R_{\mathrm{cross}}$ may become as consequential as the propagation-side term isolated here.

\section{Conclusion}\label{sec:conclusion}

This paper addresses the missing statistical interface between complex dispersive-medium characterization and clutter analysis in single-snapshot FDA-MIMO-GPR. The central contribution is a propagation-side statistical chain that maps random perturbations of the relaxation spectrum to complex permittivity, complex wavenumber, steering-vector perturbation, medium-induced clutter covariance, and total clutter covariance. Within this formulation, the total clutter covariance is written as $\bm R_c=\bm R_0+\bm R_{\mathrm{med}}$, and $\bm R_{\mathrm{med}}$ admits a KL-based modal interpretation that connects the spectrum of the underlying medium random field to the spectral structure of the clutter covariance. This formulation in turn provides an explicit route for analyzing how medium uncertainty affects effective rank, effective clutter-subspace dimension, and target--clutter separability.

The numerical evidence supports the internal consistency and practical relevance of this framework within its intended regime. The literature-informed dielectric families establish physically traceable prior scenarios, the degeneration checks confirm correct reduction to simpler limiting cases, the Monte Carlo studies show numerical closure of the main propagation chain, gprMax-based full-wave FDTD snapshots provide an independent solver-level consistency check at the covariance-spectrum level, and the validity study identifies a weak perturbation regime together with a bounded extension into a moderate regime. In the representative whitening-and-detection benchmark, the structured covariance model raises AUC from $0.593$ for the diagonal baseline to $0.753$, while prior-mismatch experiments indicate gradual rather than abrupt degradation.

The conclusion that follows from these results is therefore specific rather than universal. The present study does not establish a prior-free or fully general clutter model; instead, it shows that propagation-side medium uncertainty can be embedded into FDA-MIMO-GPR clutter covariance analysis in an explicit, interpretable, and numerically verifiable way. In that sense, the main value of the framework lies in linking physically meaningful medium variability to covariance broadening and subspace expansion through a tractable analytical construction.

The present evidence also remains bounded by clear assumptions. Quantitative accuracy is tied to the weak phase-perturbation regime, the propagation-dominated approximation excludes the coupled case in which the same medium fluctuation perturbs both propagation and local scattering strength, and the processor remains prior dependent. In addition, the validation is entirely numerical and literature informed, while gprMax-based full-wave FDTD snapshots provide a solver-level consistency check, no measured FDA-MIMO-GPR data is used. The most immediate next steps are therefore to test the framework under richer scattering and stronger heterogeneity, to extend the covariance model to include the scattering-side term $\bm R_{\beta}$ and cross term $\bm R_{\mathrm{cross}}$, and ultimately to assess the theory on measured datasets, building on the full-wave evidence already established here.

\appendices

\section{First-Order Propagation from Relaxation-Spectrum Perturbation to Steering Perturbation}\label{app:propagation}

This appendix supplements Section~\ref{subsec:random_to_steering} by making explicit the first-order propagation.

From \eqref{eq:epsilon_c} and \eqref{eq:tilde_g}, the perturbation of the complex permittivity is
\begin{equation}
	\delta\epsilon_c(\omega)
	=
	\int_{-\infty}^{\infty}
	\frac{\delta\tilde g(u)}{1+j\omega e^u}\,du
\end{equation}
Hence, $\delta\epsilon_c(\omega)$ is the image of $\delta\tilde g(u)$ under a linear Debye-type integral operator.

Under the small-perturbation condition
\begin{equation}
	|\delta\epsilon_c(\omega)|\ll |\bar{\epsilon}_c(\omega)|
\end{equation}
a first-order Taylor expansion of \eqref{eq:k_c} at $\bar{\epsilon}_c(\omega)$ yields
\begin{equation}
	\delta k_c(\omega;\bm\mu)
	\approx
	\left.
	\frac{\partial k_c(\omega;\bm\mu)}{\partial \epsilon_c(\omega)}
	\right|_{\bar{\epsilon}_c(\omega)}
	\delta\epsilon_c(\omega)
	=
	\frac{\omega\mu_0}{2\sqrt{\mu_0\bar{\epsilon}_c(\omega)}}
	\delta\epsilon_c(\omega)
\end{equation}

For the $m$th channel, by \eqref{eq:steering} and \eqref{eq:a0m},
\begin{equation}
	a_m(\theta,r;\bm\mu)
	=
	a_{0,m}(\theta,r)\exp\!\big(-j\,\delta k_c(\omega_m;\bm\mu)L_m(\theta,r)\big)
\end{equation}
If $|\delta k_c(\omega_m;\bm\mu)L_m(\theta,r)|\ll 1$, then
\begin{equation}
	\begin{aligned}
		\delta a_m(\theta,r)
		 & \triangleq
		a_m(\theta,r;\bm\mu)-a_{0,m}(\theta,r) \\
		 & \approx
		-j\,L_m(\theta,r)a_{0,m}(\theta,r)\delta k_c(\omega_m;\bm\mu)
	\end{aligned}
\end{equation}
Substituting the expression of $\delta k_c(\omega_m;\bm\mu)$ gives
\begin{equation}
	\delta a_m(\theta,r)
	\approx
	\int_{-\infty}^{\infty}
	H_m(\theta,r;u)\,\delta\tilde g(u)\,du
\end{equation}
namely,
\begin{equation}
	\delta\bm a(\theta,r)
	=
	\int_{-\infty}^{\infty}
	\bm H(\theta,r;u)\,\delta\tilde g(u)\,du
\end{equation}
This establishes the first-order operator chain used in the main text.

\section{Covariance Decomposition and a Lower Bound Linking \texorpdfstring{$r_{\mathrm{eff}}$}{reff} and \texorpdfstring{$p_\rho$}{prho}}\label{app:bridge}

This appendix supplements Sections~\ref{subsec:covariance_decomposition} and \ref{subsec:subspace_separability}.

From \eqref{eq:delta_a_vec}, the local steering covariance is
\begin{equation}
	\bm R_a(\theta,r)
	=
	\mathbb E\!\left[\delta\bm a(\theta,r)\delta\bm a(\theta,r)^H\right]
\end{equation}
Let
\begin{equation}
	\bm a(\theta,r;\bm\mu)=\bm a_0(\theta,r)+\delta\bm a(\theta,r)
\end{equation}
Then the clutter snapshot can be decomposed as
\begin{equation}
	\bm x_c=\bm x_0+\bm x_{\mathrm{med}}
\end{equation}
with
\begin{equation}
	\begin{gathered}
		\bm x_0=\iint \beta(\theta,r)\bm a_0(\theta,r)\,d\theta\,dr \\
		\bm x_{\mathrm{med}}=\iint \beta(\theta,r)\delta\bm a(\theta,r)\,d\theta\,dr
	\end{gathered}
\end{equation}
Under Assumptions~\ref{asm:second_order}--\ref{asm:propagation_dominated}, the main propagation-dominated model gives
\begin{equation}
	\bm R_c
	=
	\mathbb E[\bm x_c\bm x_c^H]
	=
	\bm R_0+\bm R_{\mathrm{med}}
\end{equation}
where
\begin{equation}
	\bm R_0
	=
	\iint
	\sigma_\beta^2(\theta,r)\,
	\bm a_0(\theta,r)\bm a_0(\theta,r)^H
	\,d\theta\,dr
\end{equation}
and
\begin{equation}
	\bm R_{\mathrm{med}}
	\equiv
	\bm R_{\mathrm{prop}}
	=
	\iint
	\sigma_\beta^2(\theta,r)\,
	\bm R_a(\theta,r)
	\,d\theta\,dr
\end{equation}
If the scattering coefficient is also perturbed by the same medium randomness, the more general first-order covariance contains the additional terms $\bm R_{\beta}$ and $\bm R_{\mathrm{cross}}$ discussed in Section~\ref{subsec:propagation_dominated_scope}. These terms are excluded from this appendix because the spectral arguments below concern the propagation-induced covariance model used in the main text.

To connect the spectral spreading of $\bm R_c$ to the effective clutter-subspace dimension, let
\begin{equation}
	p_i=\frac{\lambda_i^{(c)}}{\sum_{m=1}^{M}\lambda_m^{(c)}},
	\qquad
	\sum_{i=1}^{M}p_i=1
\end{equation}
Then
\begin{equation}
	r_{\mathrm{eff}}(\bm R_c)=\frac{1}{\sum_{i=1}^{M}p_i^2}
\end{equation}
By the definition of $p_\rho$, if $m=p_\rho$, then
\begin{equation}
	\sum_{i=1}^{m}p_i\ge \rho
\end{equation}
Using Cauchy--Schwarz,
\begin{equation}
	\left(\sum_{i=1}^{m}p_i\right)^2
	\le
	m\sum_{i=1}^{m}p_i^2
	\le
	m\sum_{i=1}^{M}p_i^2
\end{equation}
which gives
\begin{equation}
	p_\rho \ge \rho^2 r_{\mathrm{eff}}(\bm R_c)
\end{equation}
Therefore, an increase in $r_{\mathrm{eff}}(\bm R_c)$ necessarily pushes upward the lower bound of the effective clutter-subspace dimension.

\bibliography{ref}
\bibliographystyle{IEEEtran}

\end{document}